\documentclass[12pt,preprint]{aastex}

\begin{document}

\title{Ices in the edge-on disk CRBR 2422.8-3423: Spitzer spectroscopy and Monte Carlo radiative transfer modeling}

\author{Klaus M. Pontoppidan}
\affil{Leiden Observatory, P.O.Box 9513, NL-2300 RA Leiden, The Netherlands}
\email{pontoppi@strw.leidenuniv.nl}

\author{Cornelis P. Dullemond}
\affil{Max-Planck-Institut fŸr Astrophysik, P.O. Box 1317, D-85741 Garching, Germany}
\email{dullemon@mpa-garching.mpg.de}

\author{Ewine F. van Dishoeck}
\affil{Leiden Observatory, P.O.Box 9513, NL-2300 RA Leiden, The Netherlands}
\email{ewine@strw.leidenuniv.nl}

\author{Geoffrey A. Blake}
\affil{California Institute of Technology, Division of Geological and Planetary Sciences, MS 150-21, Pasadena, CA 91125}
\email{gab@gps.caltech.edu}

\author{Adwin C. A. Boogert}
\affil{California Institute of Technology, Division of Physics, Mathematics, and Astronomy, Mail Stop 105-24, Pasadena, CA 91125}
\email{acab@astro.caltech.edu}

\author{Neal J. Evans II}
\affil{Department of Astronomy, University of Texas at Austin, 1 University Station C1400, Austin, TX 78712-0259}
\email{nje@astro.as.utexas.edu}

\author{Jacqueline E. Kessler-Silacci}
\affil{Department of Astronomy, University of Texas at Austin, 1 University Station C1400, Austin, TX 78712-0259}
\email{jes@astro.as.utexas.edu}

\author{Fred Lahuis\altaffilmark{1}}
\affil{Leiden Observatory, P.O.Box 9513, NL-2300 RA Leiden, The Netherlands}
\email{freddy@sron.rug.nl}

\altaffiltext{1}{SRON National Institute for Space Research, PO Box 800, 9700 AV Groningen, The Netherlands}

   \begin{abstract}
   We present 5.2-37.2\,$\mu$m spectroscopy of the edge-on circumstellar disk CRBR 2422.8-3423 obtained using the InfraRed Spectrograph (IRS) of the Spitzer Space Telescope. The IRS spectrum is combined with ground-based 3-5\,$\mu$m spectroscopy to obtain a complete inventory of solid state material present along the line of sight toward the source.  Archival $JHK_s$ imaging as well as 350\,$\mu$m CSO mapping, 850\,$\mu$m SCUBA mapping and 3\,mm OVRO interferometry is used to obtain a set of spectro-photometric data covering 1.2--3000\,$\mu$m. The ices observed toward CRBR 2422.8-3423 are compared to archival ISOCAM-CVF 5-16\,$\mu$m ice spectra of other nearby sources within 2\arcmin. We model the object with a 2D axisymmetric (effectively 3D) Monte Carlo radiative transfer code using all the available observations to constrain the source geometry and dust composition. In particular, the location of the observed ices in the disk and envelope material is included in the model. It is found that the model disk, assuming a standard flaring structure, is too warm to contain the very large observed column density of pure CO ice, but is possibly responsible for up to 50\% of the water, CO$_2$ and minor ice species. In particular the 6.85\,$\mu$m band, tentatively due to NH$_4^+$, exhibits a prominent red wing, indicating a significant contribution from warm ice in the disk. The shape of the CO$_2$ bending mode suggests an interaction with up to 20\% of the CO ice. It is argued that the pure CO ice is located in the dense core Oph-F in front of the source seen in the submillimeter imaging, with the CO gas in the core highly depleted. 
Up to 50\% of the CO ice embedded in water or CO$_2$ ice (no more than 20\% of the total amount of CO) may still be located in the disk, assuming constant abundances of these types of CO ice throughout the system.  Discrepancies among the strength of different water ice bands are discussed. Specifically, the observed water ice libration band located at 11--13\,$\mu$m is significantly weaker than that of the model. The model is used to predict which circumstances are most favourable for direct observations of ices in edge-on circumstellar disks. Ice bands will in general be deepest for inclinations similar to the disk opening angle, i.e.
$\sim 70\degr$, except for very tenuous disks. Due to the high optical depths of typical disk mid-planes, ice absorption bands will often probe warmer ice located in the upper layers of nearly edge-on disks. The ratios between different ice bands are found to vary by up to an order of magnitude depending on disk inclination due to radiative transfer effects caused by the 2D structure of the disk. Ratios between ice bands of the same species can therefore be used to constrain the location of the ices in a circumstellar disk. 
   \end{abstract}

\section{Introduction}
It has been realized in the last decade that circumstellar disks surrounding young low-mass stars
are a common occurrence. Along with the discovery of a large number of extra-solar planets orbiting
older stars, significant observational evidence indicates that planets and planetesimals are forming in circumstellar disks within 10$^6$\,years. Understanding the chemical structure and evolution of the molecular material of circumstellar disks is crucial because it determines
the initial conditions for the composition of planets. This is especially important for the study of icy bodies
in the Solar system, such as comets and Kuiper belt objects. 

Many recent modeling and observational results indicate that the abundances of molecular gases in circumstellar disks around young low-mass stars are significantly lower than in general molecular clouds, including species such as CO, HCN and HCO$^+$ \citep{Dutrey97, Zadelhoff01, Thi04}. Although both photodissociation in the surface layer and freeze-out onto dust grains may contribute to the depletion, it is believed that freeze-out dominates in the disk mid-plane. Some
chemical models also suggest that a unique ice chemistry, different from that of dark molecular clouds,
may exist in circumstellar disks \citep{Aikawa97}. In particular, these authors predict a large abundance of CO$_2$ ice in typical circumstellar disks.

Directly observing ices in circumstellar disks is challenging. The primary reason is that cold ices can only be observed via absorption bands in the mid-infrared wavelength regime and therefore require the presence of a bright infrared source behind the icy material. The best candidates for
observations of ices in disks are circumstellar disks viewed close to edge-on, where the sight-line toward the inner bright regions passes through the disk mid-plane, which presumably is abundant in ice. A possible
exception to this may be a binary young star, in which the sight-line toward one component passes
through a disk surrounding the other. This may be the case for the T Tau binary \citep{Beck01,Hogerheijde97}, but binarity in this scenario may significantly complicate the
structure of disks in the system through dynamical and radiative interaction. Circumstellar disks
viewed at very high inclination ($\gtrsim 75\degr$)
become very faint at mid-infrared wavelengths due to the large optical depth through the disk mid-plane.
The effect this may have on observable ice absorption bands is not clear, but observations of ices in this case may not probe deeply into the disk mid-plane. The sensitivities of ground-based 8m class telescopes and the Spitzer Space Telescope are now high enough to directly observe ices in a range of Solar system-sized highly inclined disks as faint as 10\,mJy in the mid-infrared. 

In this article we present a 5.2-37.2\,$\mu$m spectrum of the edge-on disk CRBR 2422.8-3423 obtained with the Spitzer Space Telescope \citep{Werner04}. 
CRBR 2422.8-3423 ($\alpha=$16$^h$27$^m$28\fs8, $\delta=$-24\degr41\arcmin03\arcsec [J2000]) is a low luminosity source ($\sim 1\,L_{\odot}$) located in the direction of the dense core Rho Oph-F in the Ophiuchus molecular cloud complex \citep{Motte98}.
The core contains a number of other young stars as well as two prestellar condensations. The brightest submillimeter point source in the core is IRS 43, located {34\arcsec} to the north-east of CRBR 2422.8-3423.
Near-infrared imaging has shown CRBR 2422.8-3423 to be a compact (1.5\arcsec) bipolar nebula, characteristic of a disk viewed close to edge-on. The south-eastern lobe is a factor of 11 times fainter than the north-western lobe in the $K$-band  \citep{Brandner00}, showing that the disk has an inclination of $\sim70\degr$ \citep{Thi02}. The disk is unusually bright in the mid-infrared for an edge-on disk, making it an excellent target for 
infrared spectroscopy of ices.

Previous studies of ices in disks have focused on the absorption band of solid CO at 4.67\,$\mu$m observed toward isolated disks. These include Elias 18 \citep{Shuping01}, L 1489 IRS \citep{Boogert02} and CRBR 2422.8-3423 \citep{Thi02}. Of these sources, the most inclined disk is probably that of CRBR 2422.8-3423, as evidenced in the near-infrared imaging of \cite{Brandner00}. 
Spitzer spectra have also been presented toward L1489 IRS and DG Tau B \citep{Watson04}. Common for all sources is the very red (increasing flux with wavelength) near-infrared colors indicating the presence of significant amounts of foreground or envelope material; an isolated edge-on disk is expected to have near-infrared colors that are much less red due to the dominance of scattering. The CO ice profiles along the three lines of sight are quite different, 
the first two being broad, indicating that much of the CO is embedded in a water-rich ice, while
CRBR 2422.8-3423 is dominated by very deep absorption from largely pure CO ice \citep{Pontoppidan03a}. 
So far, only circumstantial evidence has been presented arguing for the location of the CO ice in the disks rather than in envelope/foreground
material.
A key point in the observational study of ices in circumstellar disks is therefore to separate the contribution to the total ice bands of the disk from the contribution of any cold foreground or remnant envelope material. A detailed study of the location of ices toward the young star Elias 29 (located
only a few arcminutes from CRBR 2422.8-3423) found that most of the absorbing material along this particular line of sight is located in foreground clouds \citep{Boogert02b}. This study provides an interesting reference to
the case of CRBR 2422.8-3423. First, Elias 29 is viewed face-on in contrast to CRBR 2422.8-3423. Second, it is has a bolometric luminosity of 36\,$L_{\odot}$, a factor of 20-30 higher than that of CRBR 2422.8-3423. \cite{Boogert02b} find from the shallow depth of the CO ice band toward Elias 29 that the average depletion of gas-phase CO along this line of sight is only a few percent. 

This article will show that detailed radiative transfer modeling of high resolution near-infrared and/or millimeter imaging combined
with high-quality mid-infrared spectroscopy can yield significant new information on the location 
and environment of ices. We use CRBR 2422.8-3423 as a case study, but emphasize that the method
can easily be applied to other $\sim$edge-on disks, using sensitive Spitzer-IRS spectroscopy. This article is organized as follows: In \S\ref{Observations_crbr}, the observational data of CRBR 2422.8-3423 and the surrounding cloud are presented. \S\ref{Inventory} discusses the solid state absorption features detected in the infrared spectra.
The radiative transfer model used for CRBR 2422.8-3423 is presented in \S\ref{Radtrans}. \S\ref{icesindisk} discusses the distribution
of ices in the CRBR 2422.8-3423 disk and \S\ref{Others}
summarizes predictions of the model regarding observations of ices in other edge-on disks.

\section{Observations}
\label{Observations_crbr}
Observations of CRBR 2422.8-3423 at wavelengths from 1.2\,$\mu$m to 3\,mm have been assembled
using both new and archival data. 
$JHK_s$ imaging by the Infrared Spectrometer and Array Camera (ISAAC) mounted on UT1-Antu of the Very Large Telescope (VLT) has been extracted from the public ESO archive\footnote{The VLT-ISAAC images were obtained at Paranal, Chile as part of the observing program 63.I-0691(A).}. The ISAAC images were first presented by \cite{Brandner00}.  $L$- and $M$-band
spectroscopy has been obtained using the Long Wavelength spectroscopic mode of ISAAC; the $M$-band spectrum was discussed in \citep{Thi02} \footnote{The VLT-ISAAC spectroscopy was obtained at Paranal, Chile as part of the observing programs 164.I-0605(A) and 71.C-0338(A)}, but we present a new observation of the $^{13}$CO ice band centered at 4.78\,$\mu$m. Spitzer 5.2-37.2\,$\mu$m spectroscopy was taken as part of the `From Molecular Cores to Planet-forming Disks' (c2d) legacy program \citep{Evans03}. A  350\,$\mu$m image was obtained with the Submillimeter High Angular Resolution Camera II (SHARC-II) at the Caltech Submillimeter Observatory (CSO), and a SCUBA 850\,$\mu$m map obtained with the James Clerk Maxwell Telescope (JCMT) was extracted from the `COMPLETE survey of star-forming regions'\footnote{See http://cfa-www.harvard.edu/COMPLETE/}. Finally, a 5-16\,$\mu$m spectral image obtained with the set of Circular Variable Filters on the Infrared Space Observatory (ISOCAM-CVF)  was extracted from the public ISO archive\footnote{In part based on observations with ISO, an ESA project with instruments funded  by ESA Member States (especially the PI countries: France, Germany, the  Netherlands and the United Kingdom.) and with the participation of ISAS  and NASA.}. The spectral image contains a number of sources in the Oph-F region and provides a useful comparison for the ice bands observed toward CRBR 2422.8-3423.

\subsection{Mid-infrared spectroscopy}
5.2-37.2\,$\mu$m spectroscopy was obtained with the Infrared Spectrograph (IRS) aboard the Spitzer Space Telescope
on March 25, 2004 using the short-low (SL) module in the 5.2-14.5\,$\mu$m range with a spectral resolving power of $R=\lambda/\Delta\lambda\sim 100$, the short-high (SH) module at 9.9-19.6\,$\mu$m and the long-high (LH) module at 18.7-37.2\,$\mu$m with a resolving power of $\sim 600$. The archival AOR key is 0009346048 for PROGID 172. Total exposure times were 28\,s, 244\,s and 484\,s in the SL, SH and LH modules, respectively. The spectra were extracted from the
pipeline (version S9.5.0) images using our c2d reduction software. The short-low spectra were extracted from pipeline products that were not corrected for stray light. The background in the SL modules was determined by fitting a Gaussian with an additive bias to the spectrum at each point along the dispersion direction. Special care was taken to match the orders of the
SH and LH echelle spectra and remove the many bad pixels of the LH module. The SH module was scaled to the SL module in their overlap region, and the LH module was then scaled to the SH module. The final spectrum was smoothed with a 2-pixel
box-car function. It should be stressed, however, that the present level of data reduction is likely to be improved significantly in the
near future, especially in terms of the shape of solid state features.

The $L$-band spectrum was obtained with ISAAC on the night of May 2, 2002 using the low-resolution
mode and the 0\farcs6 slit, yielding a resolving power of $R=600$. The $R=5\,000$ $M$-band setting centered on the 4.67\,$\mu$m band of solid CO is dicussed in \cite{Thi02}. An additional $M$-band setting was obtained on May 8, 2003 of the 4.78\,$\mu$m band of solid $^{13}$CO
with the 0\farcs3 slit yielding $R=10\,000$. The ISAAC spectra were reduced following the procedure described in \cite{Pontoppidan03a}.

The ISOCAM-CVF image (AOR key 29601813) of the Oph-F core was reduced using the CAM Interactive Analysis (CIA) package ver. DEC01
\footnote{The ISOCAM data presented in this article were analysed using ``CIA", a joint development by the ESA Astrophysics division and the ISOCAM Consortium. The ISOCAM Consortium is led by the ISOCAM PI. C. Cesarsky.}. 
A spectrum with $R=35$ was extracted from the 3$\times$3\,pixels=$18\time18\arcsec$ centered on IRS 43, the nearest bright source to CRBR 2422.8-3423.
The CVF data were first presented by \cite{Alexander03}.
\subsection{(Sub)millimeter imaging}

\begin{figure}
  \plottwo{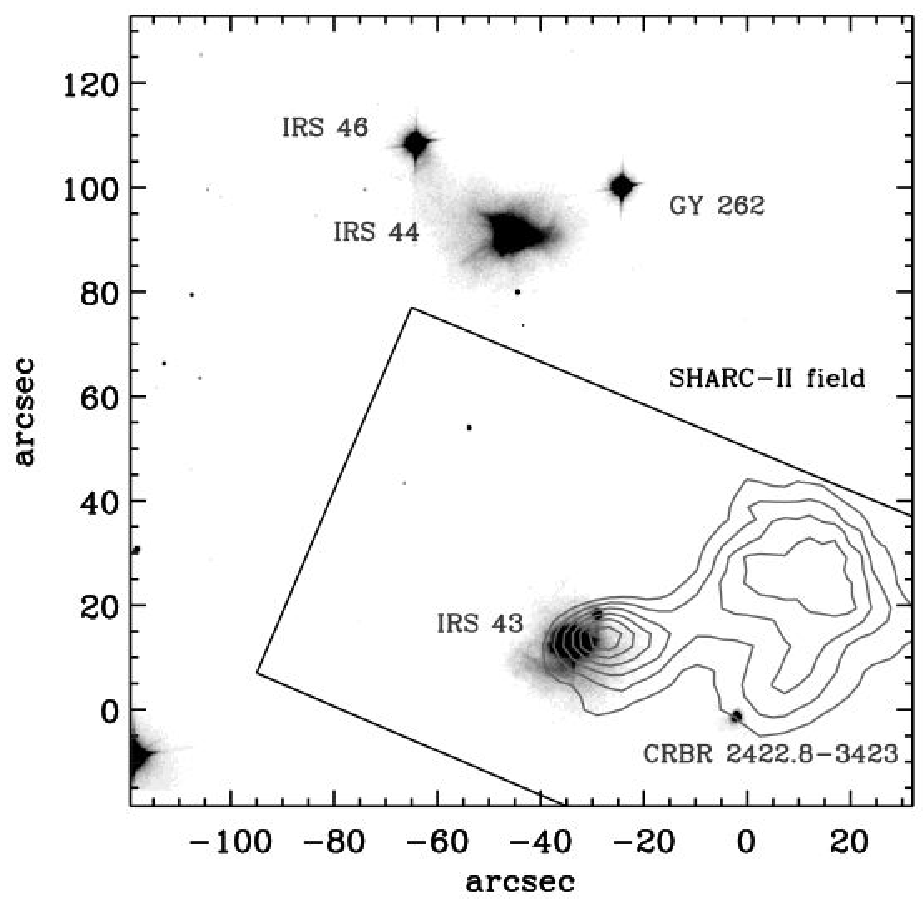}{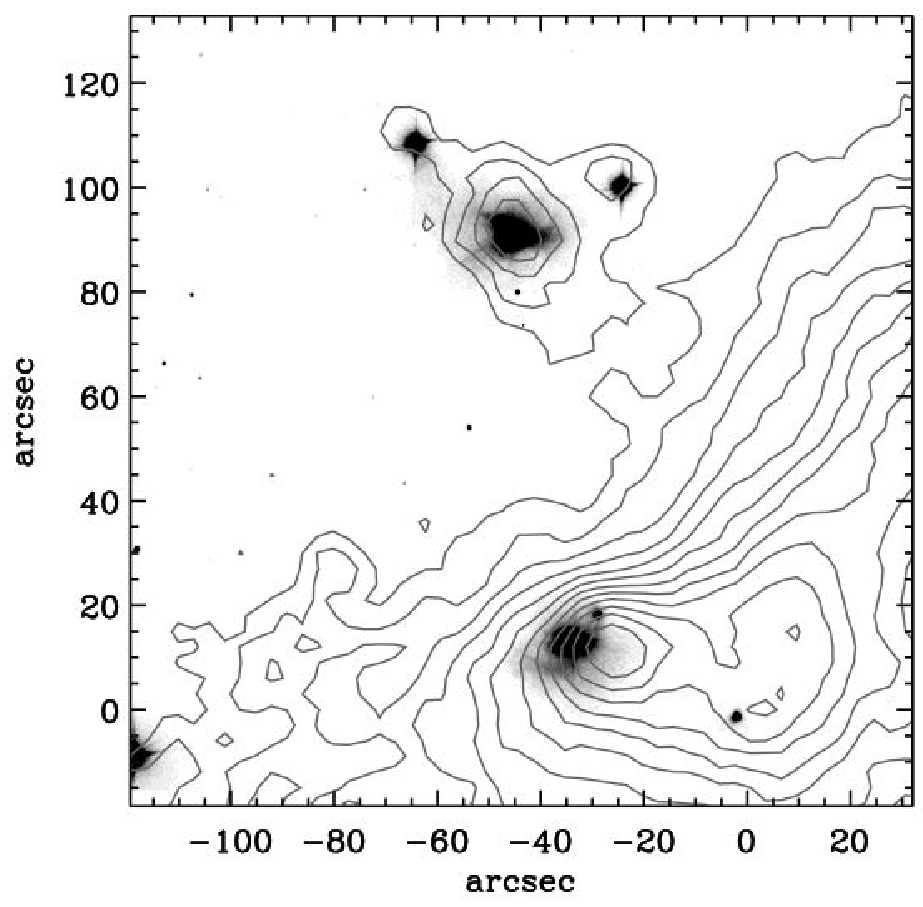}
  \caption{{\it Left panel}: SHARC II 350\,$\mu$m map of the CRBR 2422.8-3423 region overlaid on the ISAAC $K_s$-band image. The box indicates the area observed at 350\,$\mu$m. Contours start at 1.0\,Jy\,beam$^{-1}$ and increase in steps of 2.0\,Jy\,beam$^{-1}$. {\it Right panel}: SCUBA 850\,$\mu$m map from the COMPLETE survey. Contours start at 0.05\,Jy\,beam$^{-1}$ and increase in steps of 0.05\,Jy\,beam$^{-1}$. North is up and east is to the left.}
  \label{COMPLETE}
\end{figure}

The SHARC-II 350\,$\mu$m and SCUBA 850\,$\mu$m maps are shown in Fig. \ref{COMPLETE}, overplotted on the VLT-ISAAC $K$-band image of the region. It is evident from the thermal dust emission that CRBR 2422.8-3423 is situated close to several
embedded sources, at least in projection. IRS 43 is the brightest submillimeter source in the map, although the emission peak is shifted to the west by almost $10\arcsec$ in both maps. A significant
extended source corresponding to the prestellar core Oph-F MM2 \citep{Motte98}, is evident to the north-west of CRBR 2422.8-3423. No obvious submillimeter point source is associated with CRBR 2422.8-3423, although some emission is present slightly offset to the NW. It is a priori not clear whether the observed dust emission is due to emission from the edge-on disk, blended with extended emission or due to remnant envelope material surrounding CRBR 2422.8-3423. 
The offset emission from IRS 43 and CRBR 2422.8-3423 could indicate pointing errors. However, the offsets are similar in both the CSO and SCUBA maps. Additionally, IRS 44, IRS 46 and GY 262 have 850\,$\mu$m point source emission, whose positions closely coincide with the infrared sources (Fig. \ref{COMPLETE}). 

The disk is not detected at 3\,mm with the Owens Valley millimeter interferometer with a 3$\sigma$ upper limit of 3\,mJy ($8.5\arcsec\times 4.0\arcsec$ synthesized beam). 
This upper limit to the amount of compact millimeter emission from the region is sufficient to indicate that the submillimeter emission is
largely due to the core Oph-F MM2 together with possible extended remnant envelope material and not directly related to the disk.

\clearpage

\clearpage

\section{Inventory of ices}
\label{Inventory}
\begin{figure}
  \plotone{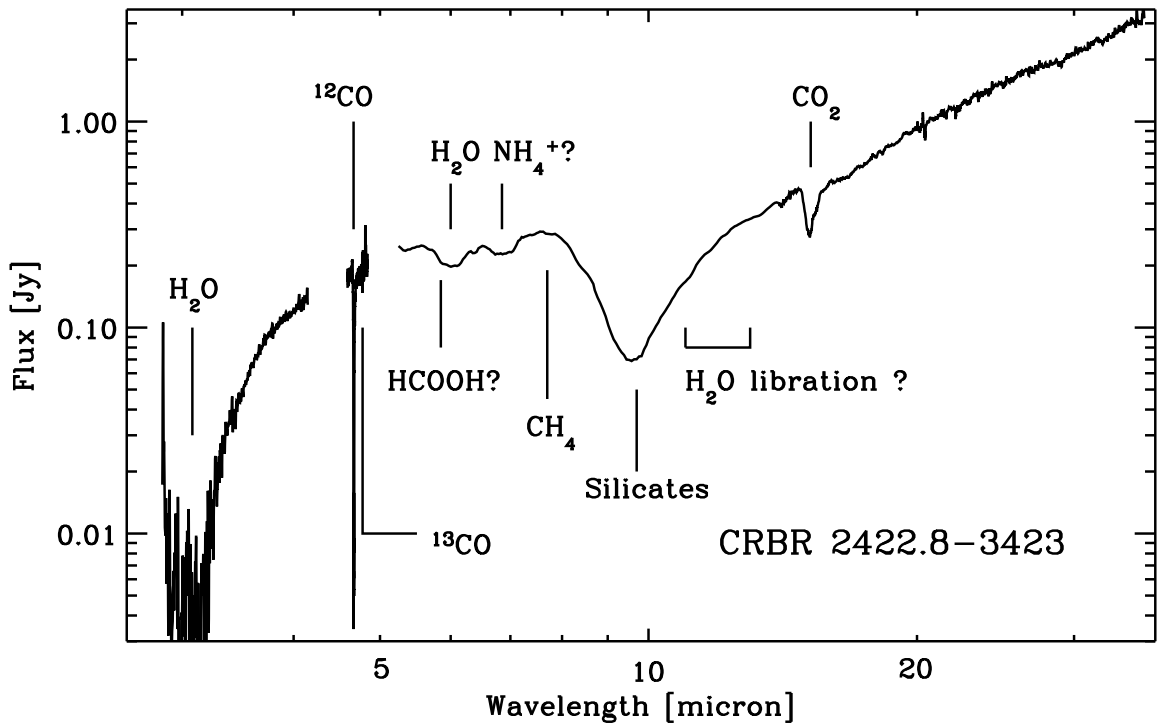}
  \caption{Combined VLT-ISAAC and Spitzer-IRS spectrum of CRBR 2422.8-3423. Detected solid state bands are marked. The water ice libration band at 11-13\,$\mu$m is largely missing (see text).}
  \label{allspec}
\end{figure}

The full 2.85--37.2\,$\mu$m spectrum toward CRBR 2422.8-3423 is shown in Fig. \ref{allspec}. There are two holes in the spectral coverage at 4.15--4.55\,$\mu$m and 4.9--5.2\,$\mu$m. The first interval contains the CO$_2$ stretching mode, which probably has a depth comparable to that of the CO stretching band at 4.67\,$\mu$m toward this source. The second interval is mostly observable from the ground, but it was not included in the ISAAC observations and is not expected to contain any strong solid state features. 

The spectrum is dominated by absorption bands due to silicates and ices along the line of sight. The strongest bands are those due to CO, H$_2$O and CO$_2$, but several weaker bands are also detected.
Specifically, the 6.85\,$\mu$m band, which still lacks an unambiguous identification, is prominent. In the following, 
the mid-infrared spectral features are discussed in order of increasing wavelength.
Table \ref{abundances} summarizes the abundances of the different ice species.

\clearpage
\clearpage

\subsection{\boldmath{$L$}-band (2.85--4.15\,\boldmath{$\mu$}m)}
The $L$-band spectrum shows a moderately deep 3.08\,$\mu$m water ice band. The band is not saturated and therefore has a well-determined optical depth of $2.3\pm 0.3$. The 3.54\,$\mu$m band due to the C--H stretch of solid methanol (CH$_3$OH) is not detected with an upper limit to the optical depth of 0.05, corresponding to an abundance relative to water ice of at most 10\%. Similarly, the unidentified `3.47\,$\mu$m feature', possibly due to NH$_3$--H$_2$O complexes \citep{Dartois02}, is not detected. 
However, the limits on the abundances are not very strict owing to the moderate water ice column density and the low flux levels of the $L$-band spectrum.
Finally, there is no detection of hydrogen recombination lines such as Br$\alpha$ at 4.07\,$\mu$m, possibly due to veiling by a strong near-infrared excess.  

\subsection{\boldmath{$M$}-band (4.55--4.90\,\boldmath{$\mu$}m)}
\label{Mband}
\begin{figure}
\includegraphics[angle=90,width=15cm]{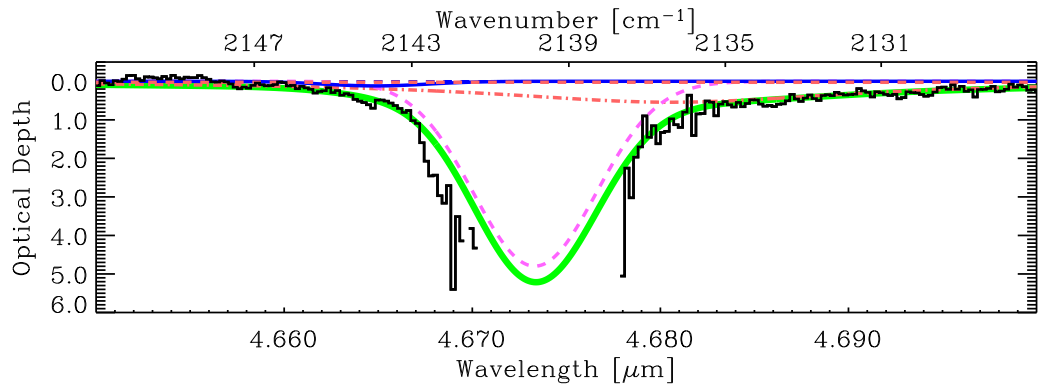}
  \caption{The $^{12}$CO ice band observed toward CRBR 2422.8-3423 with VLT-ISAAC. The
  curves indicate the three-component decomposition of \cite{Pontoppidan03a}.}
  \label{12CO}
\end{figure}

The $M$-band spectrum shows an extremely deep CO ice band at 4.67\,$\mu$m as well as
narrow absorption lines from low-J rovibrational transitions of gaseous CO \citep{Thi02,Pontoppidan03a}. In general, the shape of the main $^{12}$CO ice band observed in dense clouds can be decomposed into three components with a fundamental set of centers and widths \citep{Pontoppidan03a}, such that all observed $^{12}$CO ice bands can be well-fitted by only varying the relative intensities of the bands. The three-component decomposition is shown in Fig. \ref{12CO}. The CO ice band is dominated by the `middle' component at 2139.9\,cm$^{-1}$, likely due to mostly pure CO. The `red'
CO ice component at 2136.5\,cm$^{-1}$, probably due to CO in a water-rich environment, is detected with an optical depth of 10\% of that of pure CO. The contribution of the `blue' component centered on 2143.7\,cm$^{-1}$ is negligible toward CRBR 2422.8-3423. This component is not unambiguously identified, but has been associated with CO mixed with CO$_2$ \citep{Boogert02} or with the longitudinal optical mode of crystalline CO, which appears when the background source is polarized \citep{Pontoppidan03a}. The dominance of the 2139.9\,cm$^{-1}$ component corresponds to about 80\% of the CO molecules being in a mostly apolar environment. 

\begin{figure}
 \plotone{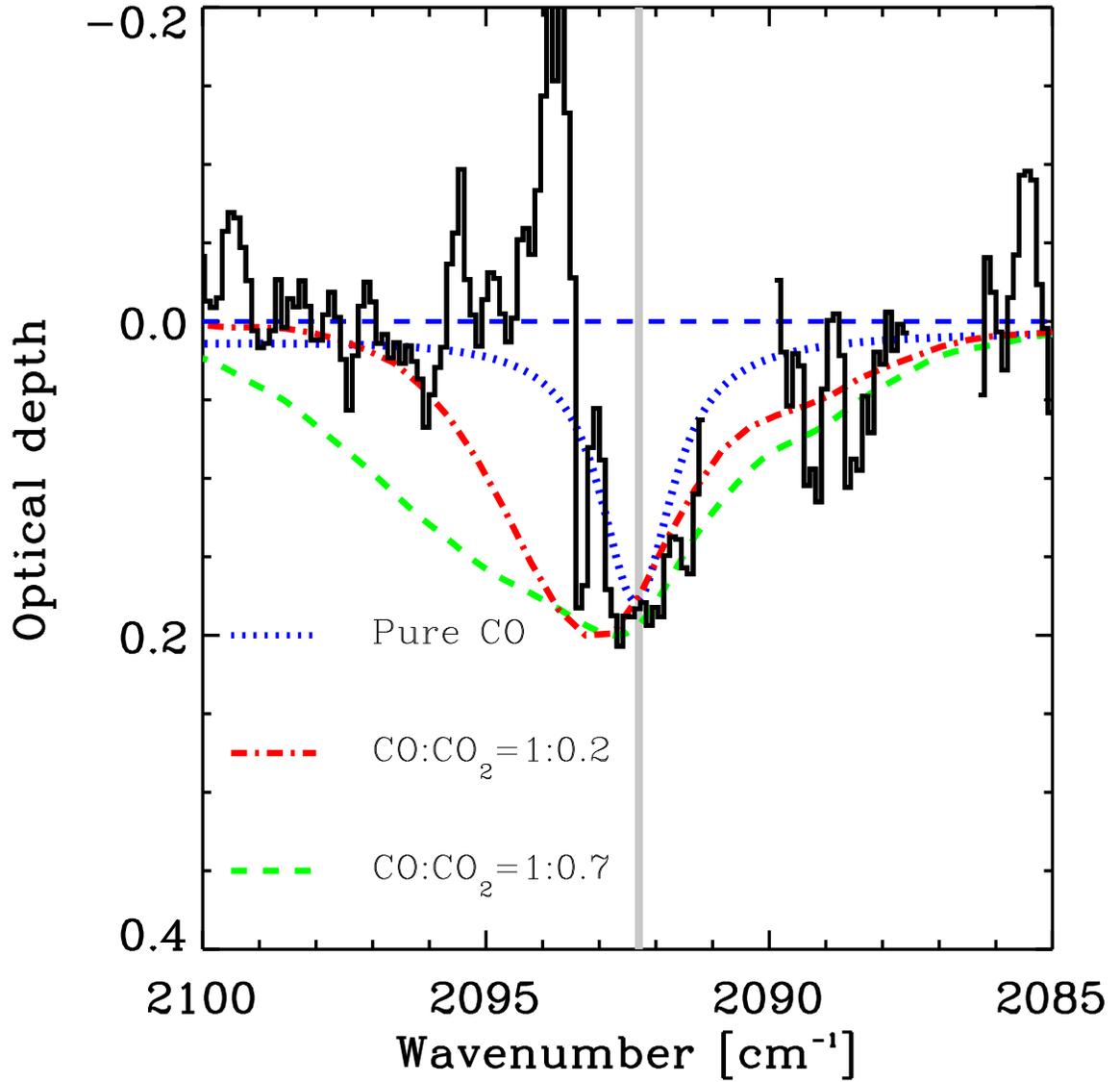}
  \caption{The $^{13}$CO ice band observed toward CRBR 2422.8-3423 with VLT-ISAAC. The
  dotted curve is pure CO at 10\,K, the dashed and dashed-dotted curves are CO:CO$_2$=1:0.7 and CO:CO$_2$=1:0.21, respectively, both at 30\,K. The difference in temperature of the laboratory spectra is not significant. Regions strongly affected by
  residuals from telluric lines have been removed from the spectrum. }
  \label{13CO}
\end{figure}

The $^{13}$CO band at 4.78\,$\mu$m is clearly detected with an optical depth of $0.17\pm0.03$. This is only the third object for which $^{13}$CO ice is found, and clearly testifies to the exceptionally large solid CO column density. In Fig. \ref{13CO}, the band is compared with laboratory spectroscopy from the Leiden database \citep{Ehrenfreund96}. The $^{13}$CO band is useful for determining the molecular environment of the CO ice. This is due to the insensitivity of the $^{13}$CO band to the grain shape effects that dominate that for $^{12}$CO. Pure CO produces a narrow, Lorentzian $^{13}$CO band centered on 2092\,cm$^{-1}$, while relatively small contamination (at least 10-20\%) from other molecules tends to broaden the band significantly. The presence of telluric CO gas phase lines on either side of the $^{13}$CO ice band, however, complicates the profile analysis. Visual inspection of Fig. \ref{13CO} indicates that the $^{13}$CO band may be slightly broader than what is expected for pure CO.  Assuming pure CO results in a $^{12}$CO/$^{13}$CO ratio of $45\pm15$. This is somewhat lower than the value of 70 derived for other sources \citep{Boogert02, Pontoppidan03a}.  
A broader $^{13}$CO ice band will result in a larger column density of $^{13}$CO and therefore an even
lower isotopic ratio. 

\clearpage

\clearpage

\subsection{The 5--12\,\boldmath{$\mu$}m region}
\label{5to8}
The short-low region of the Spitzer spectrum, covering 5.2-14.5\,$\mu$m, contains clear detections of the enigmatic 6.0/6.85\,$\mu$m solid-state band complex as well as the 9.7\,$\mu$m silicate band. The silicate absorption band has an optical depth of 1.4-1.8, depending on how the 
continuum is defined, giving a small $\tau_{\rm H_2O}/\tau_{\rm Silicate}$ ratio of $\sim$1.4. This is often seen for embedded sources and may indicate that the silicate band has been partially filled in by emission \citep{Alexander03}. 

\begin{figure}
\includegraphics[width=10cm]{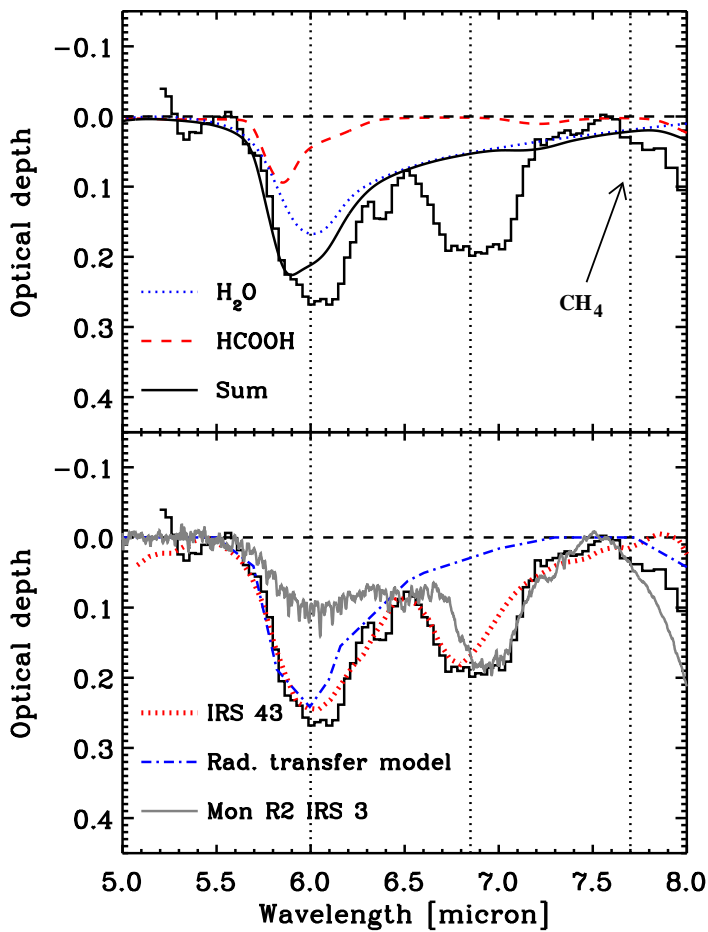}
  \caption{The 5--8\,$\mu$m region observed toward CRBR 2422.8-3423. {\it Upper panel}: The observed spectrum is compared to a laboratory spectrum of water ice at 10\,K scaled to an optical depth of 2.3 for the 3.1\,$\mu$m band (dotted curve).
A laboratory spectrum of pure HCOOH at 10\,K has been added to account for the blue wing of the band (dashed curve). The solid curve is the sum of the two laboratory spectra. {\it Lower panel}: The spectrum is compared to the ISOCAM-CVF spectrum of the nearby IRS 43 (dotted curve) and the ISO-SWS spectrum of Mon R2 IRS 3 from \cite{Keane01} (solid grey curve). Note that the red shoulder of the 6.85\,$\mu$m band is absent toward IRS 43 but can be well fitted by the Mon R2 IRS 3 spectrum. Three vertical dotted lines are added to guide the eye at 6.0, 6.85\,$\mu$m and 7.7\,$\mu$m. The IRS 43 spectrum has not been scaled to match the CRBR 2422.8-3423 spectrum, but happens to have almost the same optical depth. The dot-dashed curve is the radiative transfer model described in \S\ref{Radtrans}.}
  \label{CRBR_6mu}
\end{figure}

Fig. \ref{CRBR_6mu} shows the 5-8\,$\mu$m region of the CRBR 2422.8-3423 Spitzer spectrum 
on an optical depth scale. Laboratory spectra \citep[e.g.][]{Schutte96,Schutte99} as well the ISOCAM-CVF spectrum of IRS 43 and the ISO Short Wavelength Spectrometer (SWS) spectrum of the massive young star Mon R2 IRS 3 from \cite{Keane01} are used for comparison. The continuum is a 2nd order polynomial chosen to pass through the 5.3-5.5 and the 14.8-14.9\,$\mu$m regions. There is some uncertainty involved in choosing a continuum in this way, but the shapes of the bands are not affected, even if their total optical depths may be uncertain by up to $\sim 20\%$. The spectrum shows a prominent 6.0\,$\mu$m feature, which
is partly due to water ice. A laboratory spectrum of pure water ice deposited at 10\,K and scaled to the 3.08\,$\mu$m band is compared to the observed spectrum. It is seen that the red wing of the 6.0\,$\mu$m
water band can be fitted, while the main laboratory band can account for only 60\% of the total observed band depth. This is very common and most likely indicates that other ice species contribute significantly
to the 6.0\,$\mu$m band \citep{Keane01}, although radiative transfer effects can also cause a similar effect (see \S\ref{Others}). A significant blue shoulder to the 6.0\,$\mu$m band is detected
at 5.85\,$\mu$m. This shoulder has been found in other high mass \citep{Keane01} and
low-mass \citep{Pontoppidan04} sources. Plausible candidates include both formaldehyde (H$_2$CO)
and formic acid (HCOOH). In this case, HCOOH provides the best fit, both because pure H$_2$CO
is centered at 5.82\,$\mu$m, which is too blue for the observed spectrum, and because the 3.47\,$\mu$m band of H$_2$CO is not detected. A secure identification is not possible at the present, however, thanks to signal-to-noise limitations and level of data reduction. 
The HCOOH band shown in Fig. \ref{CRBR_6mu} corresponds to an abundance relative to water ice of 2.4\%.

A narrow absorption feature is seen at 6.35\,$\mu$m. A sharp feature at this wavelength has not
been observed for other sources. The band is possibly an instrumental artifact, although no obvious problems are seen
in the pipeline spectroscopic images. Furthermore, the Spitzer Science Center pipeline extraction software ver. 9.5.0 also produces an absorption band at this wavelength comparable to that obtained with our own extraction. We therefore regard the detection of this band as tentative because it
has not previously been reported toward any other source. A possible identification can be the formate ion (HCOO$^-$), which has a strong band centered around 6.33\,$\mu$m. The presence of HCOO$^-$
is possible, given the presence of HCOOH, since it can be produced by acid-base reactions between,
for instance, HCOOH and NH$_3$. Only 0.4\% of HCOO$^-$ relative to water ice is required to produce a similar band. Another possible explanation is that an emission band due to
polycyclic aromatic hydrocarbons (PAHs) is present at 6.2\,$\mu$m, thereby creating the illusion of an
absorption band at 6.35\,$\mu$m. The spectral resolution is too low to distinguish between these possibilities.  Finally, solid O$_2$ has a very weak band at 6.45\,$\mu$m. However, the depth of the observed band requires an extremely high column density of solid O$_2$ of $>10^{19}$\,cm$^{-2}$ corresponding to an abundance of $2.5\times 10^{-3}$ for the largest band strength measured of $\rm 1\times 10^{-19}$\,cm\,molec$^{-1}$ \citep{Ehrenfreund92}. The band strength of the O$_2$ band has been measured to be 1--2 orders of magnitude smaller, depending on ice composition \citep{vandenbussche99}. 

\clearpage
\clearpage

CRBR 2422.8-3423 also shows a prominent 6.85\,$\mu$m band. The carrier of this band remains 
unknown, although a significant number of candidates have been tested. Substantial arguments have been given in favour of NH$_4^+$ \citep[][and references therein]{Schutte03}. A significant concern for a charged carrier of such a strong band is the lack of observed counter-ions, but \cite{Schutte03} argue that the counter-ions may be distributed among a range of carriers, rendering any single ionic species unobservable. \cite{Keane01} noted the presence of two components to the 6.85\,$\mu$m band through an empirical decomposition of the band for a sample of high-mass stars. In Fig. \ref{CRBR_6mu} the 5-8\,$\mu$m spectrum of CRBR 2422.8-3423 is compared to laboratory spectra as well as the ISOCAM-CVF spectrum of IRS 43 and the ISO-SWS spectrum of the massive young star Mon R2 IRS 3 from \cite{Keane01}. Mon R2 IRS 3 is included here because it has the most extreme red-shifted 6.85\,$\mu$m band observed to date. The 6.0/6.85\,$\mu$m band complexes toward CRBR 2422.8-3423 and IRS 43 have, by chance, nearly identical optical depths, while the spectrum of Mon R2 IRS 3 has been scaled by a factor of 0.7. The main difference is the complete lack of the
red shoulder of the 6.85\,$\mu$m band toward IRS 43 as compared to CRBR 2422.8-3423. This difference may be of considerable significance. The Mon R2 IRS 3 spectrum is seen to provide a very good fit to the red part of the CRBR 2422.8-3423 6.85\,$\mu$m band, which is entirely missing from the IRS 43 spectrum. A shift of the 6.85\,$\mu$m component to 7.0\,$\mu$m was suggested by \cite{Keane01} to be an effect of thermal processing, meaning that the line of sight toward CRBR 2422.8-3423 contains warm ice in addition to the cold ice probed by the CO ice band. \cite{Schutte03} have shown that NH$_4^+$ is shifted to the position of the reddest observed band toward Mon R2 IRS 3 at temperatures in excess of 200\,K. The corresponding temperature for interstellar ice is possibly somewhat lower.
The location of material at high temperatures along the line of sight toward CRBR 2422.8-3423 is discussed in terms of the radiative transfer modeling in \S\ref{icesindisk}. The abundance of NH$_4^+$ required to produce a 6.85\,$\mu$m feature like that observed toward CRBR 2422.8-3423 is 10\% relative to water ice, assuming a band strength of $4.0\times 10^{-17}\,$cm\,molec$^{-1}$ \citep{Schutte03}.

The 7.7\,$\mu$m band of solid CH$_4$ is tentatively detected, but unresolved at the Spitzer-IRS resolution. The silicate absorption band at 9.7\,$\mu$m is largely featureless (see Fig. \ref{allspec}) and does not show strong evidence
for the 9.35\,$\mu$m `umbrella' mode due to solid NH$_3$ or the 9.75\,$\mu$m CO stretching mode of solid CH$_3$OH.

\subsection{CO\boldmath{$_2$} 15.2\,\boldmath{$\mu$}m bending mode}

\begin{figure}
\includegraphics[width=10cm]{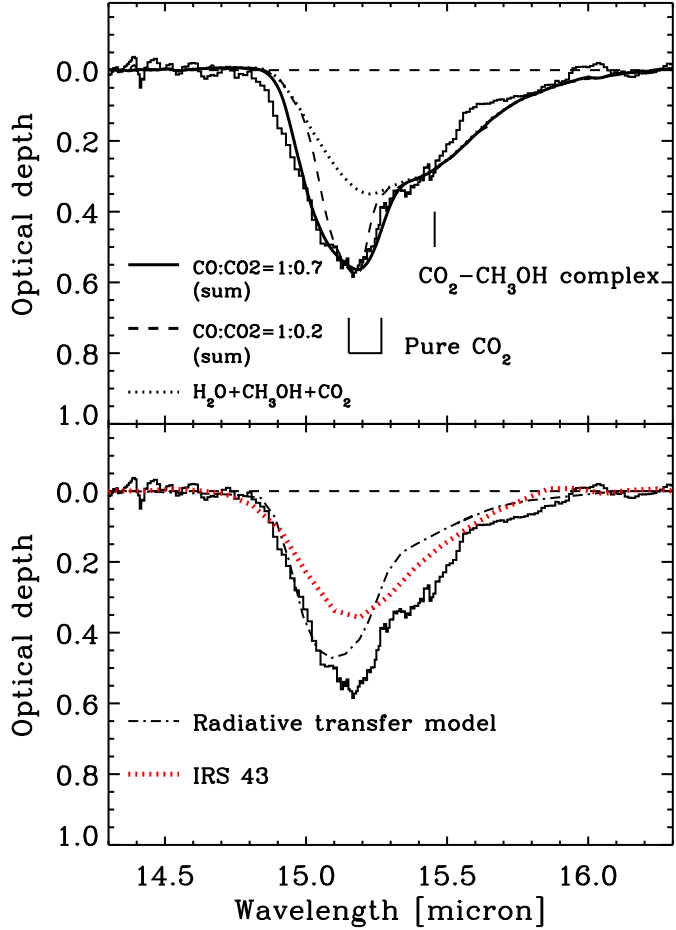}
  \caption{The 15.2\,$\mu$m CO$_2$ ice bending mode observed toward CRBR 2422.8-3422 on an optical depth scale. {\it Upper panel}: The curves show a sum of laboratory spectra of a water and methanol-rich mixture (H$_2$O:CH$_3$OH:CO$_2$=1:0.6:1 at 30\,K) and a CO rich mixture. The solid curve shows a fit using CO:CO$_2$=1:0.7 at 10\,K for the CO-rich mixture, while the dashed curve uses CO:CO$_2$=1:0.2 at 10\,K. A continuous distribution of ellipsoid (CDE) grain model has been used to convert the optical constants to absorption coefficients. {\it Lower panel}:  The dotted curve is the CO$_2$ band observed by ISOCAM-CVF toward IRS 43. The dot-dashed curve is the band extracted from our radiative transfer model (see \S\ref{Radtrans}). The spectral resolution of the ISOCAM spectrum is 0.4\,$\mu$m at the location of the CO$_2$ bending mode. }
  \label{CO2}
\end{figure}

The short-high spectrum shows a strong band at 15.2\,$\mu$m due to the bending mode of CO$_2$ ice. The CO$_2$ bending mode observed toward CRBR 2422.8-3423 is shown in Fig.~\ref{CO2} where it is compared to the laboratory spectroscopy of \cite{Ehrenfreund96} as well as the CO$_2$ bending mode observed toward IRS 43.
The depth of the band corresponds to an abundance of CO$_2$ ice relative to water ice of $N({\rm CO_2})/N({\rm H_2O})=0.32$ (see Table \ref{abundances}). This is significantly higher than the typical value of 0.2 observed for high-mass stars \citep{Gerakines99}, but is similar to
abundances recently measured toward some low-mass stars \citep{Boogert04}. Laboratory experiments have shown that the shape of the CO$_2$ bending mode is a sensitive tracer of ice composition and thermal history. In particular, substructure bands can be used to examine whether solid methanol is
mixed with the CO$_2$ and the extent of ice heating \citep{Ehrenfreund99}. The location of the CO$_2$ bending mode within the Spitzer high spectral resolution coverage enables a detailed study of the band shape.
 
\clearpage
\begin{table}
\footnotesize
\centering
\caption{Average abundances of ices observed toward CRBR 2422.8-3423}
\begin{tabular}{llllll}
\hline
\hline
& Column density&Abun. rel. to H$_2$O ice & Abun. rel. to H$_2$ gas & band strength\\
& $10^{18}\,$cm$^{-2}$ &&&\\
\hline
H$_2$O &3.6&1&$9\times 10^{-5}$&$2\times 10^{-16}\,^a$\\
CO$_2$ &1.16&0.32&$2.9\times 10^{-5}$&$1.1\times 10^{-17}\,^a$\\
$^{12}$CO pure&2.8&0.78&$7\times 10^{-5}$&$1.1\times 10^{-17}\,^a$\\
$^{12}$CO in water&0.68&0.19&$1.7\times 10^{-5}$&$1.1\times 10^{-17}\,^a$\\
NH$_4^+$$^d$&0.36&0.10&$9\times 10^{-6}$&$4.0\times 10^{-17}\,^e$\\
$^{13}$CO&0.06&0.017&$1.5\times 10^{-6}$&$1.1\times 10^{-17}\,^a$\\
CH$_4$$^d$&0.13&0.035&$3\times 10^{-6}$&$8.0\times 10^{-18}\,^b$\\
CH$_3$OH&$<0.36$&$<0.10$&$<9\times 10^{-6}$&$5.3\times 10^{-18}\,^b$\\
HCOOH$^d$&0.09&0.02&$2\times 10^{-6}$&$6.7\times 10^{-17}\,^c$\\

\hline
\end{tabular}
\begin{itemize}
\item[$^a$] \cite{Gerakines95}
\item[$^b$] \cite{Kerkhof99}
\item[$^c$] \cite{Schutte99}
\item[$^d$] Suggested identification (see text).
\item[$^e$] \cite{Schutte03}

\end{itemize}
\label{abundances}
\end{table}

\clearpage

Pure, partially crystalline CO$_2$ at 10\,K  is split into two narrow bands at 660\,cm$^{-1}$ (15.15\,$\mu$m) and 655\,cm$^{-1}$ (15.25\,$\mu$m). There is some evidence from laboratory experiments that pure, fully amorphous CO$_2$, which does not show this characteristic splitting of the bending mode, may be formed under certain conditions \citep{Falk87}. 
Toward CRBR 2422.8-3423, a strong, relatively narrow component is clearly present around 660\,cm$^{-1}$ (15.15\,$\mu$m). Inspection of Fig. \ref{CO2}
shows that a narrow band at 655\,cm$^{-1}$ is absent, ruling out significant amounts of the pure, partially crystalline CO$_2$ toward CRBR 2422.8-3423. The absence of the 655\,cm$^{-1}$ band indicates that the structure of the CO$_2$ ice is different from that generally observed toward high-mass stars \citep{Gerakines99}. In addition to the narrow component, the observed CO$_2$ band has a significant red shoulder at 647\,cm$^{-1}$ (15.45\,$\mu$m), which can possibly be assigned to a CO$_2$--CH$_3$OH complex \citep{Gerakines99}. This shoulder is most prominent for
cold ices ($<50$\,K) and when CO$_2$, H$_2$O and CH$_3$OH are present in the mixture in
similar concentrations. A very good fit is obtained with a H$_2$O:CH$_3$OH:CO$_2$=1:0.6:1 mixture.
The abundance of CH$_3$OH required by the fit is 9\% relative to water ice, which is close to, but consistent with the observed upper limit of methanol ice.

A better theoretical and experimental understanding of the conditions under
which different phases of solid CO$_2$ are formed will allow us to explore the structure of CO$_2$ ice around low-mass protostars further.  For example, an alternative way to create a single narrow absorption band of CO$_2$ is to use a mixture with another
ice species. Since a mixture with H$_2$O cannot produce the narrow blue component, the only other candidate known to be abundant in this source is CO. Indeed, a binary mixture with CO is found to yield an excellent fit to the rest of the band, and the percentage of the total column of `pure' CO required to be mixed with CO$_2$ is 20\%. 

The presence of a binary CO:CO$_2$ component is qualitatively consistent with the $^{13}$CO band.
Quantitatively, the best fit to the CO$_2$ bending mode is a mixture with CO and CO$_2$
in approximately equal concentration, while the $^{13}$CO band is best fitted by a mixture in which
CO$_2$ is much more dilute. Since the total ratio of CO to CO$_2$ along the line of sight is $\sim 3$, this fits well with a scenario in which a small part ($\sim 20\%$) of the CO is mixed with the CO$_2$, while the rest is almost pure CO. 
A binary CO:CO$_2$ mixture should produce a blue-shifted and broadened $^{12}$CO ice band,
and there is no indication of this in the observed $M$-band spectrum (see \S\ref{Mband}). Since the CO ice band toward CRBR 2422.8-3423 is saturated, the significance of this is not clear. 

In summary, the line of sight toward CRBR 2422.8-3423 shows significant evidence
for some interaction between part of the CO$_2$ and the CO ice components, while most of the
two species remain separate. Alternatively, the narrow component of the CO$_2$ bending mode may be due to pure, but {\it amorphous} CO$_2$. One may speculate that the CO:CO$_2$ ice mixture or amorphous CO$_2$ ice resides in the disk material, mostly on the basis of circumstantial evidence. First, this component seems to be much less prominent
toward IRS 43, although an observation of this line of sight at higher spectral resolution is required to confirm this. Second, a broadened $^{13}$CO band has not previously been observed
in lines of sight toward embedded sources. However, considering that CRBR 2422.8-3423 is only the third source with a $^{13}$CO detection, this can hardly be considered conclusive.

\clearpage
\clearpage

\section{2D continuum radiative transfer}
\label{Radtrans}
In order to better understand the location of ices in the disk and the foreground or envelope of CRBR 2422.8-3423, a 3D Monte Carlo radiative transfer model was constructed for an axisymmetric density structure. The aim was to fit the dust density structure
simultaneously with the high resolution spectral energy distribution and the near-infrared imaging. 
We use the RADMC code \citep[see][]{Dullemond04} to calculate the temperature structure and scattering source function
for a given density distribution. Spectral energy distributions and images are then calculated
using the ray-tracing capabilities of the code RADICAL \citep{Dullemond00}. The
Monte Carlo radiative transfer is effectively performed in full 3D, but the density structure
is required to be axisymmetric. The scattering is assumed to be isotropic, although
a full treatment of scattering is being developed. The effects from
anisotropic scattering are relatively small for wavelengths longer than 2\,$\mu$m
and grains smaller than a few $\mu$m. However, this does not necessarily rule out that the near-infrared images are 
somewhat affected by anisotropic scattering.  

\subsection{Dust model}
Because it is necessary to adjust the abundances of ice species independently, a dust model must be constructed for the radiative transfer calculation to provide opacities that can reproduce the observed spectral features. Furthermore, to adequately
fit the relatively narrow ice bands, a higher spectral resolution is required for the opacities than is available in published tables.
The adopted dust model assumes spherical silicate grains with inclusions of carbonaceous material,
surrounded by a spherical ice mantle. In the diffuse medium, the silicate and carbonaceous grains are likely in separate dust populations \citep{Draine03}. However, when grains coagulate in dense molecular clouds, these populations may be mixed. Thus for convenience, a single grain population is adopted here.   The silicate-carbon core optical constants are calculated using
Maxwell-Garnett effective medium theory \citep{BH}:

\begin{equation}
\epsilon_{\rm av}=\epsilon_{\rm mat}\left[ 1+\frac{3f(\epsilon_{\rm inc}-\epsilon_{\rm mat})/(\epsilon_{\rm inc}+2\epsilon_{\rm mat})}{1-f(\epsilon_{\rm inc}-\epsilon_{\rm mat})/(\epsilon_{\rm inc}+2\epsilon_{\rm mat})}\right],
\label{MG}
\end{equation}
where $\epsilon_{\rm mat}$ and $\epsilon_{\rm inc}$ are the dielectric functions of the matrix (silicates)
and spherical inclusions (carbon), respectively. $f$ is the volume fraction of the inclusions. The carbon component is chosen as the inclusion, since it has the smallest volume fraction.
The opacities are then obtained using the coated sphere
Mie code from \cite{BH}. A single grain size of 0.5\,$\mu$m is used, since the shape of the opacity law is mostly affected by smaller grains at
the shorter wavelengths ($<2\,\mu$m), where little information is available due to the heavy extinction
of the source. Ices are assumed to evaporate at 90\,K, except CO, which evaporates at 20\,K \citep{Sandford93}. Above 90\,K,  only the silicate/carbon cores are left. The grains are assumed to be compact. `Fluffy', coagulated grains have increased opacity at all wavelengths \citep{OH94}; consequently, we will overestimate the required dust mass by assuming compact grains.

Optical constants for the ice are taken from the Leiden database of ices \citep{Ehrenfreund97}, those for the oxygen-rich silicates from \cite{OH94}, while the carbonaceous constants are from \cite{Jaeger98} for carbon clusters formed at 800\,K. The choice of the carbonaceous component is the most uncertain, since few spectral diagnostics are available in dense clouds and embedded objects to guide the choice. The carbonaceous material created in the laboratory experiences a significant change in properties for pyrolization temperatures between 600 and 800\,K, reflecting a larger contribution from graphitic material at the higher temperature. Here, the optical constants for carbon dust created at 800\,K are chosen because this best reproduces the observed strength of the 9.7\,$\mu$m feature. The 600\,K
sample results in opacities that cause the 9.7\,$\mu$m silicate feature to be over-estimated in 
the model fit to the observed spectrum, while creating very high far-infrared to millimeter opacities. The volume
fraction of carbonaceous inclusions used is 0.15. This number is dependent on the chosen set of optical constants for the carbon dust. A lower pyrolization temperature creates a lower carbon dust opacity in the mid-infrared, causing the volume fraction of carbon dust required to fit the silicate feature to increase by up to a factor of two. The significance of the carbon volume fraction is therefore limited. For comparison, the dust model of \cite{Weingartner01} uses a carbon volume fraction of $\sim 0.3$.

It is not possible to create a database of ice optical constants that covers even a small fraction
of the parameter space of ice abundances. As a consequence,
ice abundances in the mantle are varied using Eq. \ref{MG}. As
shown in \S\ref{Inventory}, much of the CO$_2$ ice is embedded in water ice. Therefore, as a starting point an ice mixture of H$_2$O:CO$_2$:CO=1:0.2:0.03 at 10\,K is used. The CO$_2$ and CO abundance is increased by adding inclusions of CO:CO$_2$=1:0.7. Beyond 20\,$\mu$m, a revised version of the ice opacity of \cite{Warren84} for crystalline water ice is used, since no optical constants are available for amorphous water ice at these wavelengths. The CO$_2$:CO inclusions can be replaced by pure CO to test the location of this component
in the disk/envelope system (see \S\ref{icesindisk}). The water ice abundance for the ice-coated dust grains is $9\times 10^{-5}$ relative to H$_2$,
assuming an ice density of 0.6\,g\,cm$^{-3}$ and a gas-to-dust mass ratio of 100.

\begin{figure}
\plottwo{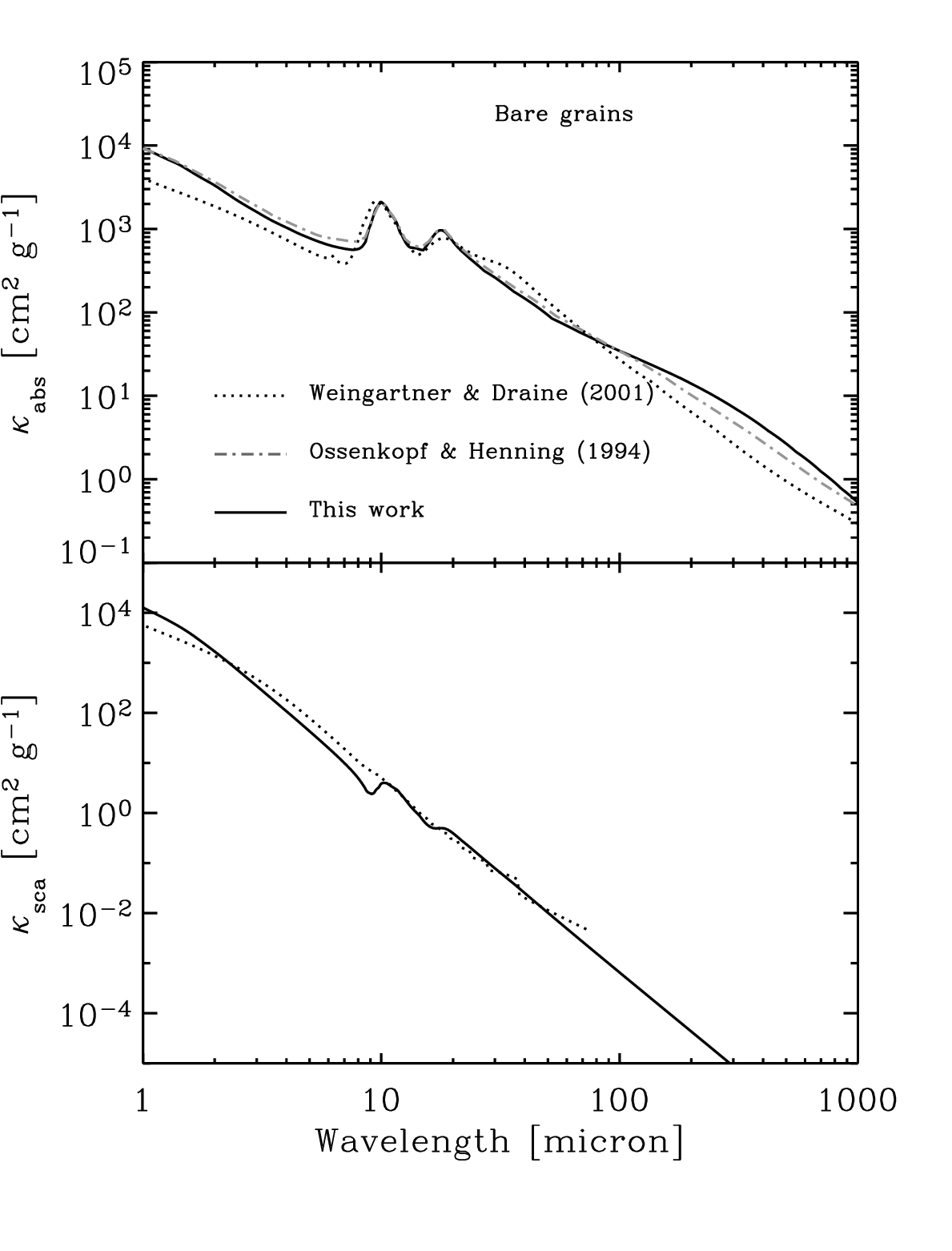}{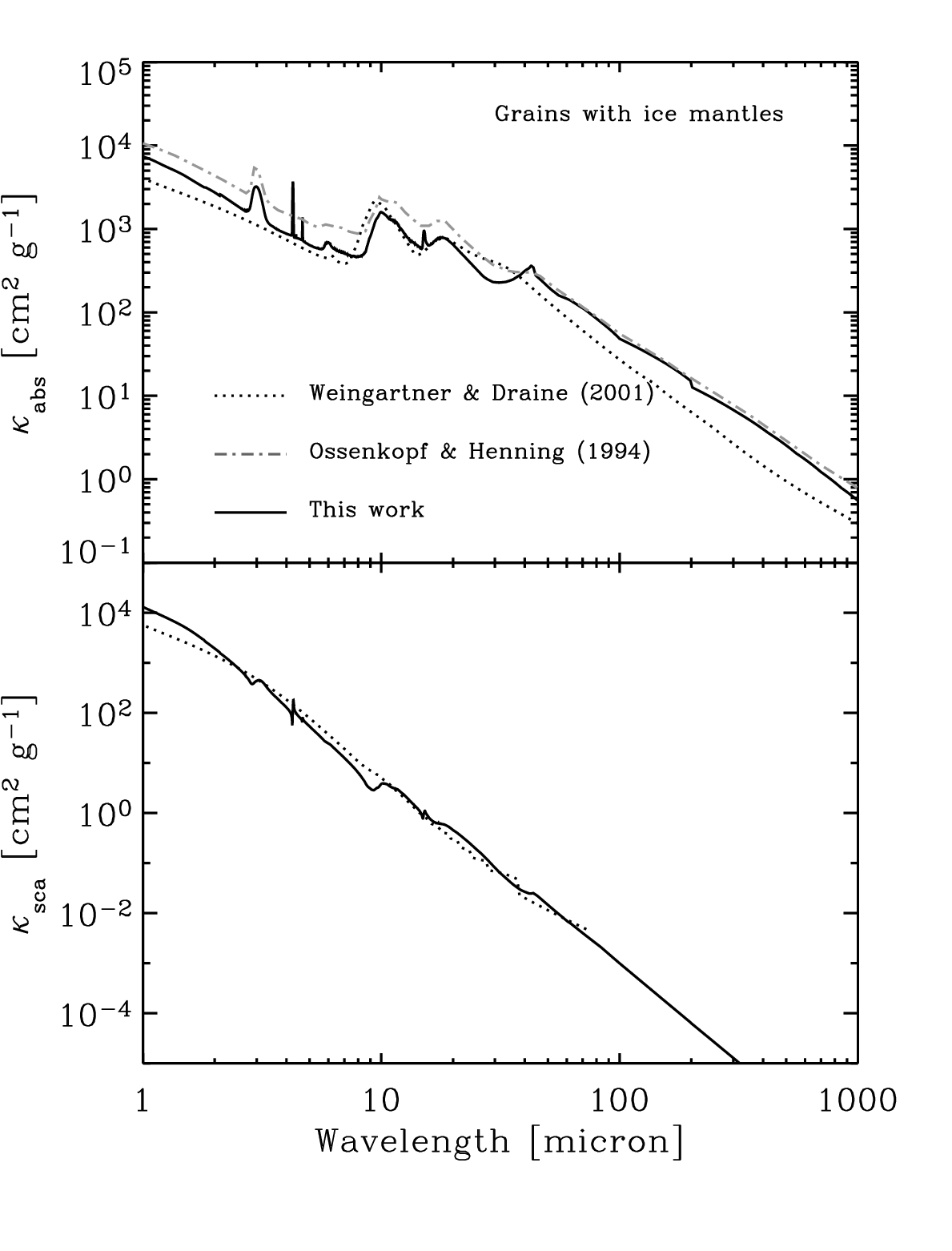}
  \caption{Examples of dust mass absorption (top) and scattering (bottom) efficiencies used as input for the radiative transfer model compared to
  opacities from the literature. {\it Left panels}: Mass absorption and scattering coefficients for dust grains without ice.
  The fairly high absorption coefficients at long wavelengths are largely due to the choice of carbonaceous
  optical constants. {\it Right panels}: Dust mass absorption and scattering coefficients for ice-coated grains with a water ice abundance of $9\times 10^{-5}$
  relative to H$_2$, assuming a gas-to-dust ratio of 100 and an ice density of 0.6\,g\,cm$^{-3}$ (porous amorphous water ice). The opacities are relative to the mass of the {\it refractory} dust mass in order to preserve the dust-to-gas mass in a model when also ice freezes out.
  The opacities are compared to those of \cite{OH94} for compact grains with and without a thin ice mantle and those of \cite{Weingartner01}
  appropriate for diffuse medium dust.}
  \label{Opacities}
\end{figure}

Fig. \ref{Opacities} compares the dust opacities (not including the gas mass) to those by \cite{OH94} and \cite{Weingartner01}. The calculated opacity matches fairly well that of \cite{OH94} for compact bare grains (the first column of their Table 1), whose grain coagulation model
produces opacities that are slightly larger at all wavelengths. The opacities show a larger variation once ice mantles are added. In particular, our higher resolution opacities include the narrow ice bands
from CO and CO$_2$ at 4.27 and 4.67\,$\mu$m. Additionally, we also find a significant contribution from the 11-13\,$\mu$m water
ice libration band in our opacity. Note also the presence of the 45\,$\mu$m band from crystalline
water ice due to the use of optical constants for higher temperature water ice at long wavelengths.
The presence of this band does not affect the continuum modeling results.  

\clearpage
\clearpage

\subsection{Disk structure}
\label{DiskStruct}
The aim of the radiative transfer model is to empirically fit the dust structure of the disk, rather than
to construct a self-consistent dynamical model. Basically, this means that a `generic' disk structure is
used in which the flaring angle is a free parameter, independent of the central object or the dust model.
The vertical density structure is assumed to be Gaussian. A Gaussian density profile is appropriate for 
a vertically isothermal disk. Note, however, that the model disk is not isothermal and no assumption is made concerning the temperature, which is calculated independently using the Monte Carlo code. Non-isothermal models have vertical density profiles that are
similar to a Gaussian, only in the uppermost layers are noticable
deviations seen \citep{Dullemond02}.
Effects such as grain settling will produce a vertical dust density profile which peaks
more strongly toward the disk mid-plane, but here we are most interested in modeling the structure of the disk where
the optical depth at mid-infrared wavelengths is moderate, away from the densest parts of the mid-plane. 
A dust density profile which is strongly peaked toward the mid-plane
(e.g. due to dust sedimentation) is not probed by the mid-infrared
spectrum as long as the wing of the profile resembles that of a Gaussian
with a larger scale height, so that the photosphere of the disk lies well
above the midplane density concentration. We therefore do not include
a midplane dust layer into our model.
The disk density is thus modeled as a function of radius, $R$ and height above the disk plane, $Z$:

\begin{equation}
\rho(R,Z) = \frac{\Sigma(R)}{H_p(R)\sqrt{2\pi}}\exp\left(-\frac{Z^2}{2H_p(R)^2}\right),
\end{equation}
where $\Sigma(R)=\Sigma_{\rm disk}\times(R/R_{\rm disk})^{-p}$ is the surface density and 

\begin{equation}
H_p(R)/R=(H_{\rm disk}/R_{\rm disk})\times(R/R_{\rm disk})^{2/7} 
\end{equation}
is the disk scale height. The flaring power of $2/7$ is that determined for a self-irradiated passive disk \citep{CG97}. The parameters varied in the model are $\Sigma_{\rm disk}$, determining the
disk mass, and $H_{\rm disk}$ determining the disk opening angle measured at the outer edge of the disk, $R_{\rm disk}$. 
The disk opening angle can be constrained with some degree of confidence by the near-infrared imaging. The surface density is assumed to vary with $p=1$. The inner edge of the disk is determined by the dust sublimation
temperature, here assumed to be 1700\,K, and a lower bound to $R_{\rm disk}$ is constrained by the $K$-band image to 90\,AU (see Fig. \ref{JHK_MODEL}). The surface density is not sharply truncated at the outer edge, but follows a steep
power law of $R^{-12}$ outside of $R_{\rm disk}$. 

The central object is
assumed to be a typical pre-main sequence star with an effective temperature of $\sim 4500\,$K. The 
radius of 2 solar radii (corresponding to a luminosity of 1.4\,$L_{\odot}$) is then constrained by the SED. These parameters correspond to a pre-main sequence star of age $3\times 10^6\,\rm years$, using the evolutionary tracks of \cite{Siess00}. The model output is not very sensitive to the exact choice of temperature of the central star, only to the total luminosity.
Specifically, the star could be cooler and therefore younger. An effective temperature of 4000\,K, for example,
corresponds to an age of $10^{6}\,$years at the luminosity of CRBR 2422.8-3423.

Foreground or envelope material is added via a spherical shell of constant density. This is necessary to produce the observed high extinction toward the source. The temperature 
and any deviation from a constant density of the foreground material is not constrained due to the lack of far-infrared photometry. The location of the foreground material is discussed in the next section.

Finally, a puffed-up inner rim is added. The rationale behind a puffed-up inner rim is that the star over-heats an inner vertical wall of dust located at the dust sublimation temperature, causing it to expand in the vertical direction \citep{Natta01}. This is modeled as a puffed-up scale height at the inner edge of the disk and a drop-off to a maximum radius where the flaring disk takes over. Specifically, the disk opening angle at the inner radius, $R_{\rm rim}$, is increased to $H_{\rm rim}$. The presence of a puffed-up inner rim makes it possible to reproduce the high mid-infrared fluxes of the source, and at the same time sufficiently extinct the central star. It is thought that a puffed-up inner rim is most prominent for disks surrounding intermediate mass stars  \citep{Natta01,Dullemond01}, but it has also recently been found to be important for T Tauri stars \citep{Muzerolle03}. The observed SED of CRBR 2422.8-3423 clearly requires significant excess infrared emission around 5\,$\mu$m, which is hard to explain without introducing an inner rim. Also, at the inclination of CRBR 2422.8-3423, direct emission from the central star tends to dominate over scattered light from the disk surface if no inner rim
is used, contrary to what is observed in the near-infrared images.

The density distribution is sampled on a polar ($R$,$\Theta$) grid with a logarithmic spacing in the $R$-direction to ensure that all structures are properly resolved. Likewise, the spectral energy distributions are calculated using rays with impact parameters that are sampled on a logarithmic grid centered on the star.
At each wavelength, the flux is determined by integrating over a predefined `observational' aperture.
The images are calculated as follows: Pixels are arranged on a regular rectangular matrix and ray-tracing along the ray belonging to each pixel is performed. However, some extra care has to be taken with this
procedure. Because the pixels are equi-distant, small scale structure in the inner parts of the disk (such as the star itself) may not be resolved. This will cause an underestimate
of the flux in the inner pixel of the grid. To ensure that the image reproduces the flux at all scales, including scales that are much smaller than a pixel, the following approach was taken: For each final model image, a set of three images was calculated on three different scales, all centered on the central star. The sampling chosen for the final image was 200 pixels
for 1000\,AU. Therefore, two images with 200 pixel sampling were calculated for the inner 10\,AU and
0.1\,AU. The smallest image was then resampled to a $2\times2$\,pixel grid, which was used to replace
the inner pixels in the 10\,AU image. This method was then repeated for the 1000\,AU image. 
The smallest image fully resolves the central star.

\subsection{Best model fit}

\clearpage
\begin{figure}
  \plottwo{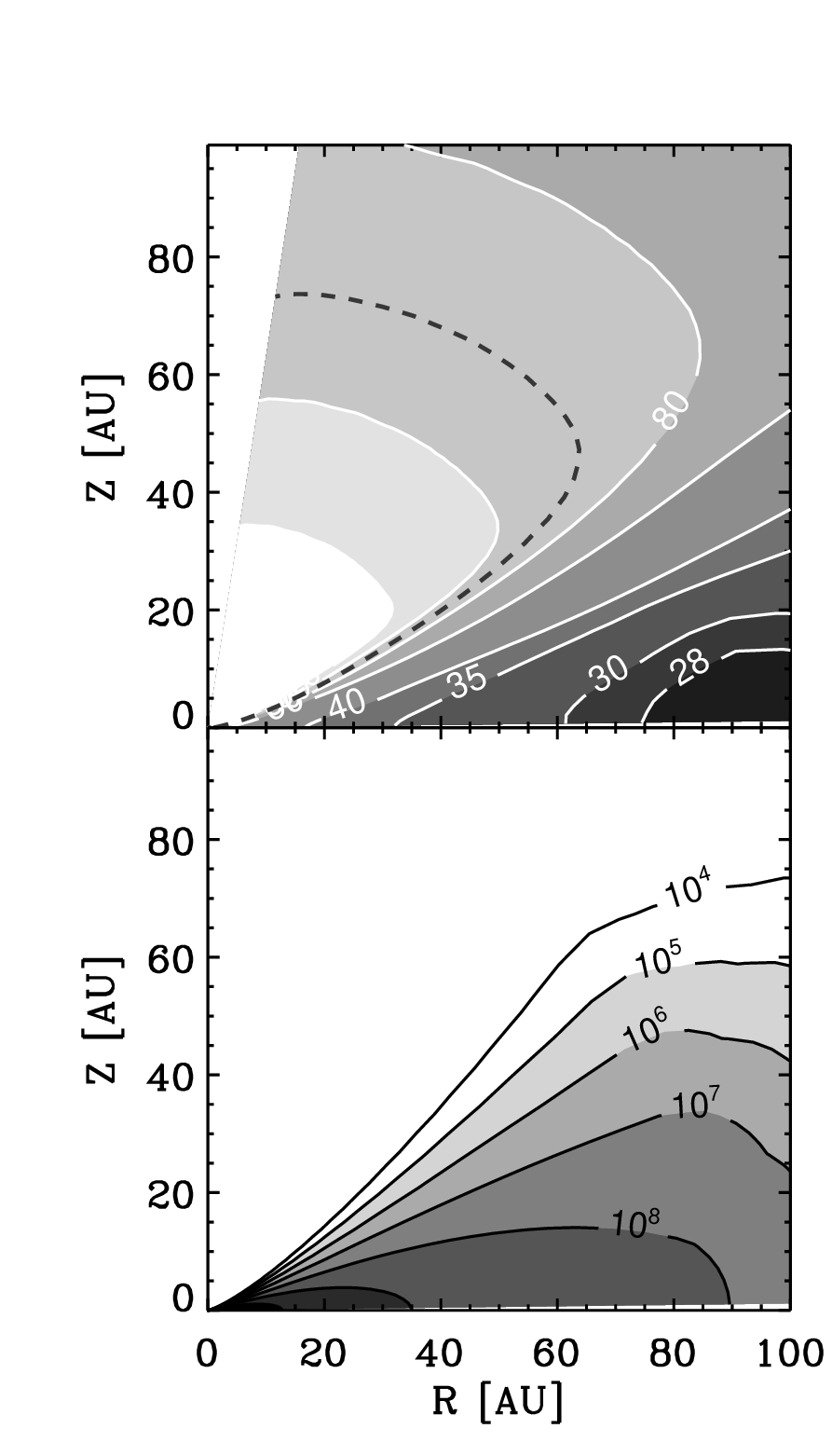}{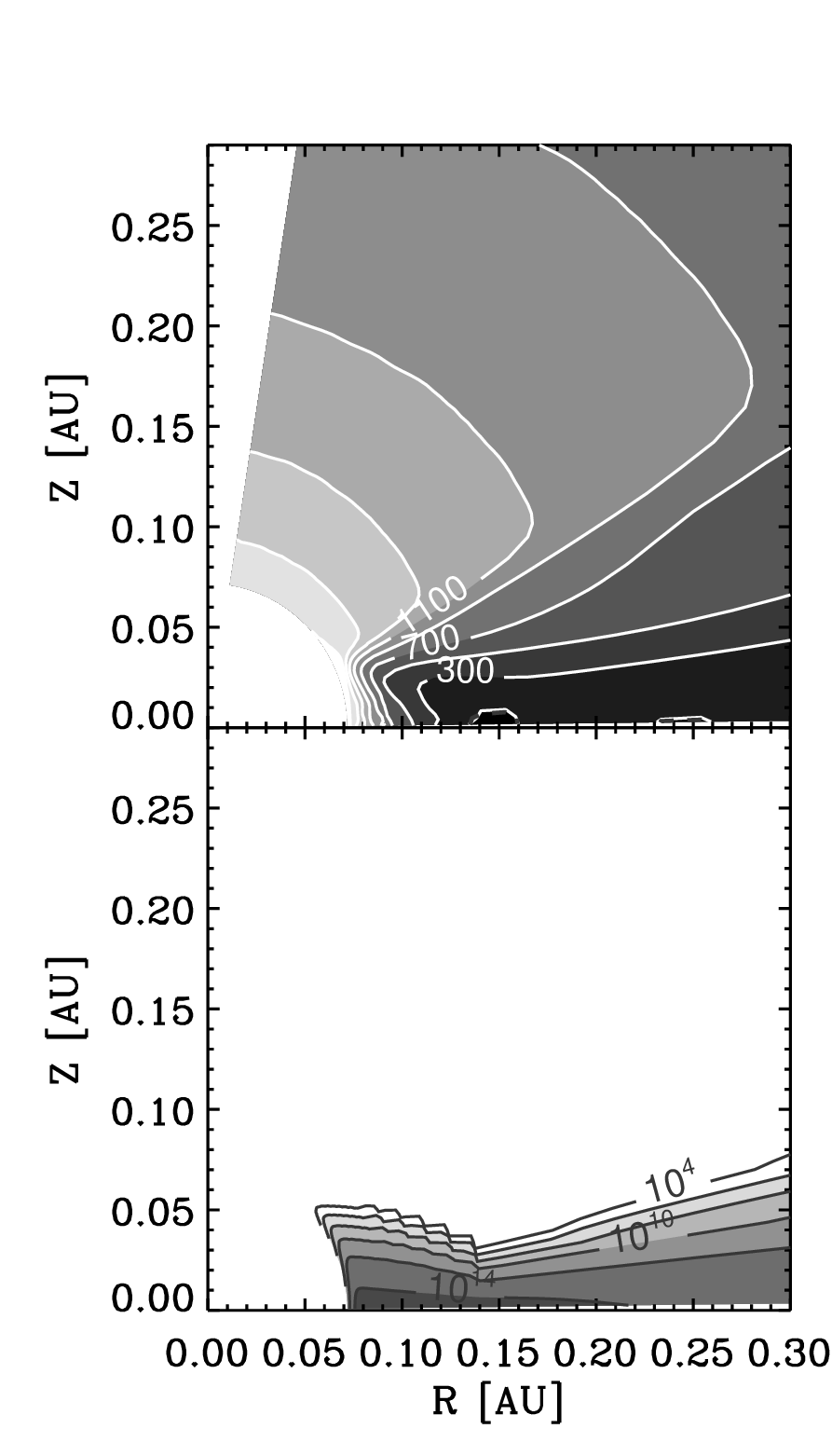}  
  \caption{Structure of the best-fitting model disk. The left panels show the entire disk, while the right panels show a zoom-in of the inner rim region. {\it Upper panels}: Dust temperature in Kelvin. The dashed curve indicates the (water) ice evaporation zone at 90\,K. {\it Lower panels}: H$_2$ density assuming a gas-to-dust ratio of 100 in units of cm$^{-3}$. Contours are shown in steps of 1 dex (left) and 2 dex (right).}
  \label{diskstruct}
\end{figure}

\begin{figure}
\epsscale{0.75}
  \plottwo{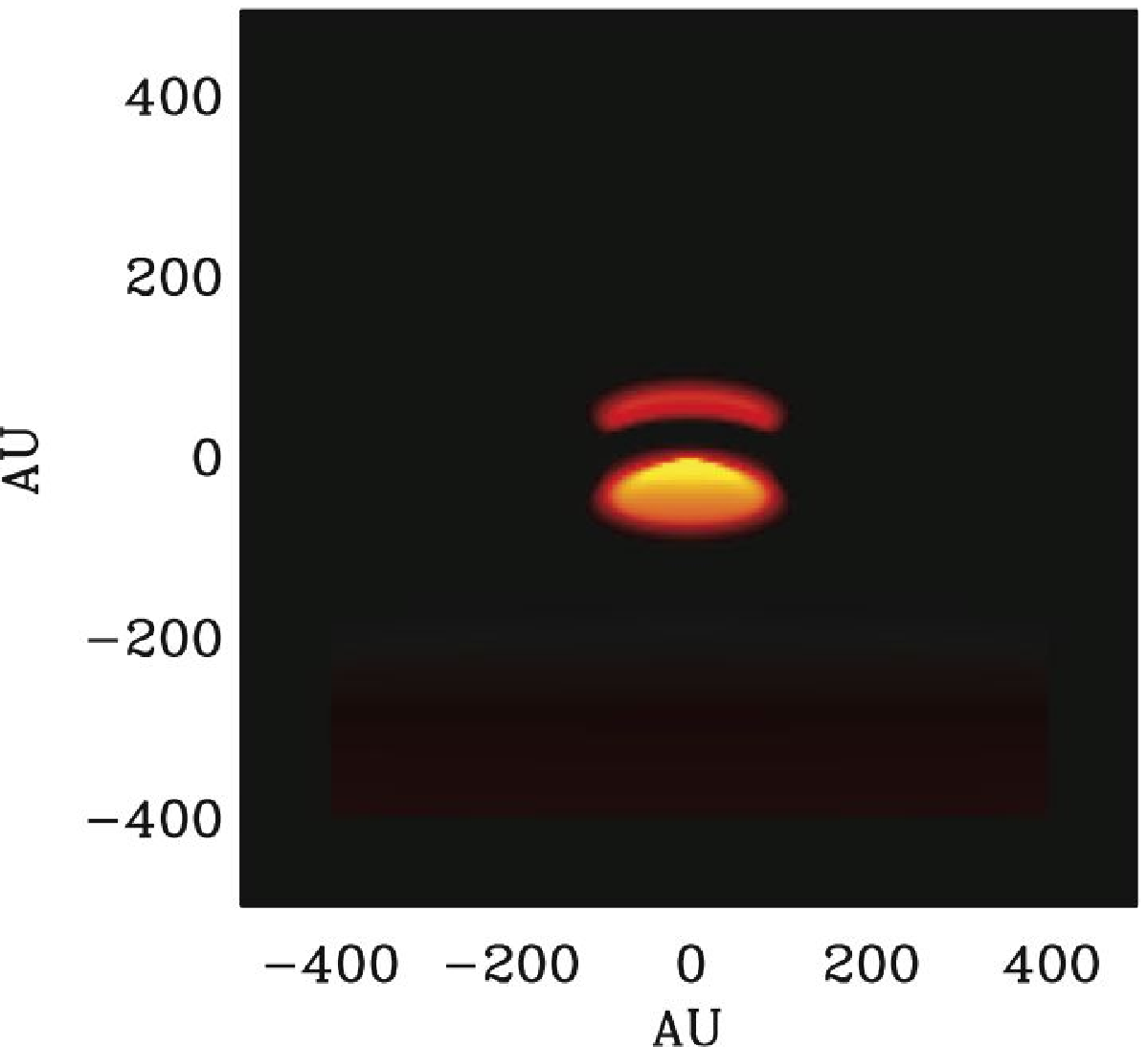}{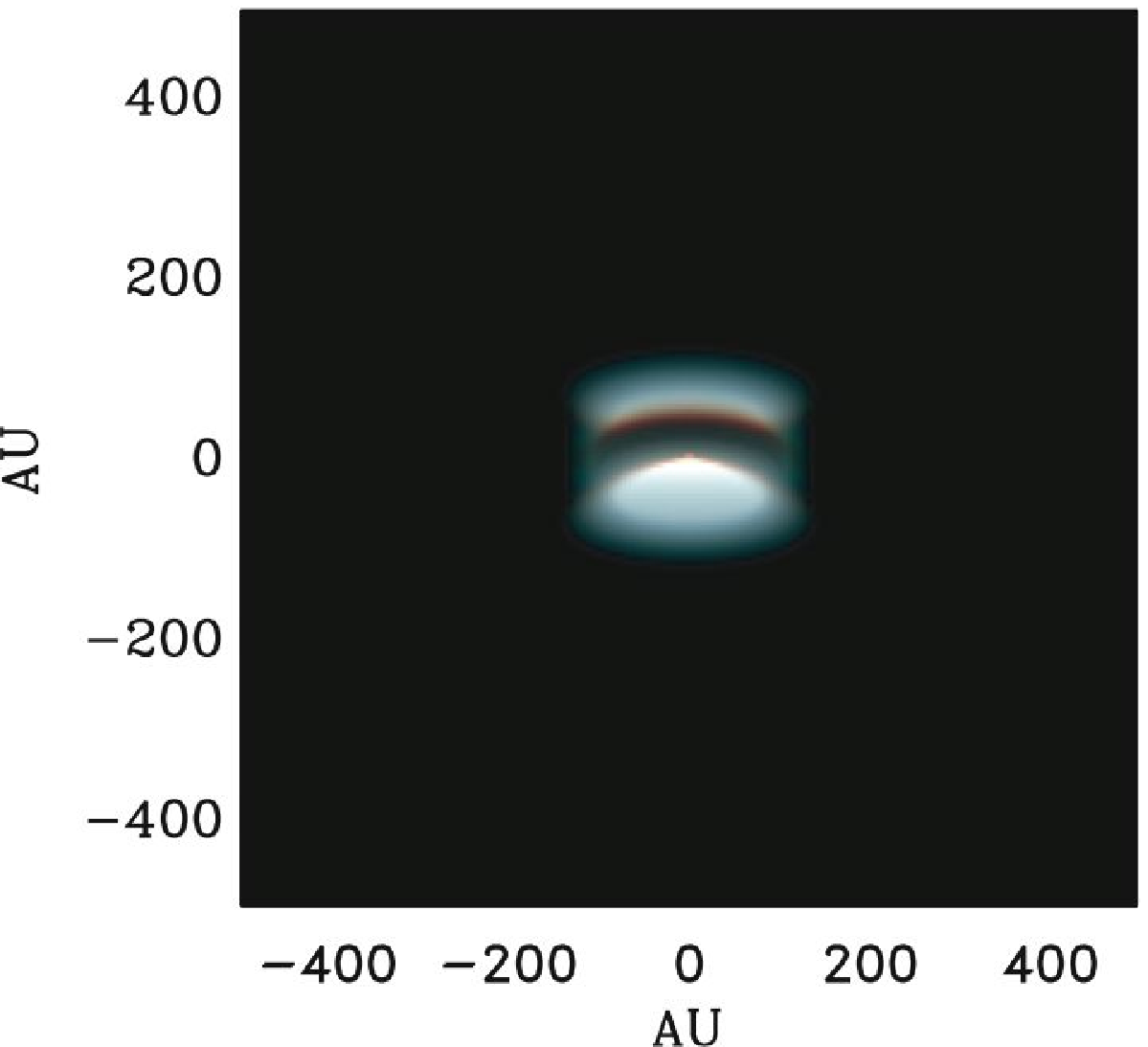}
  \plottwo{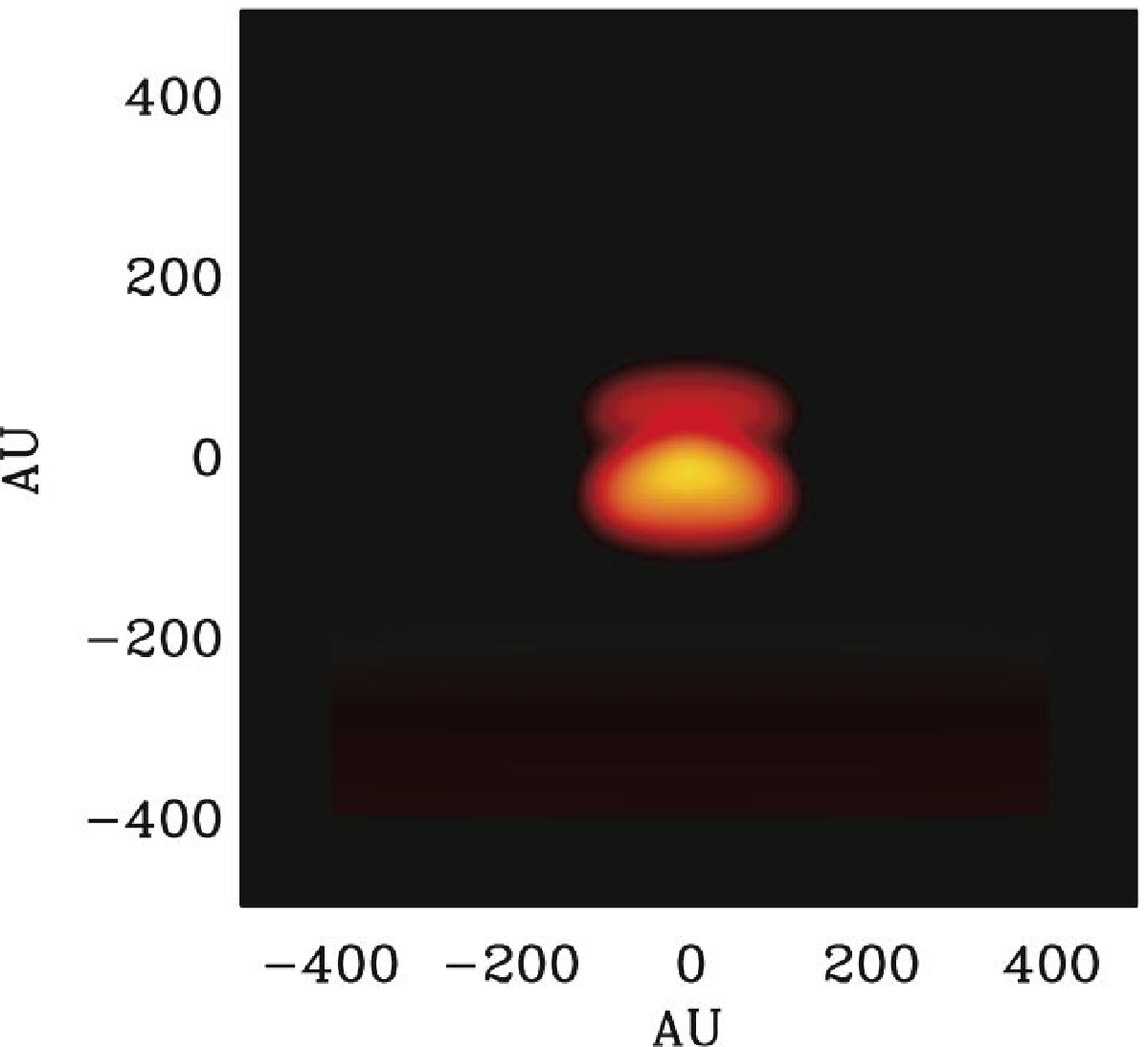}{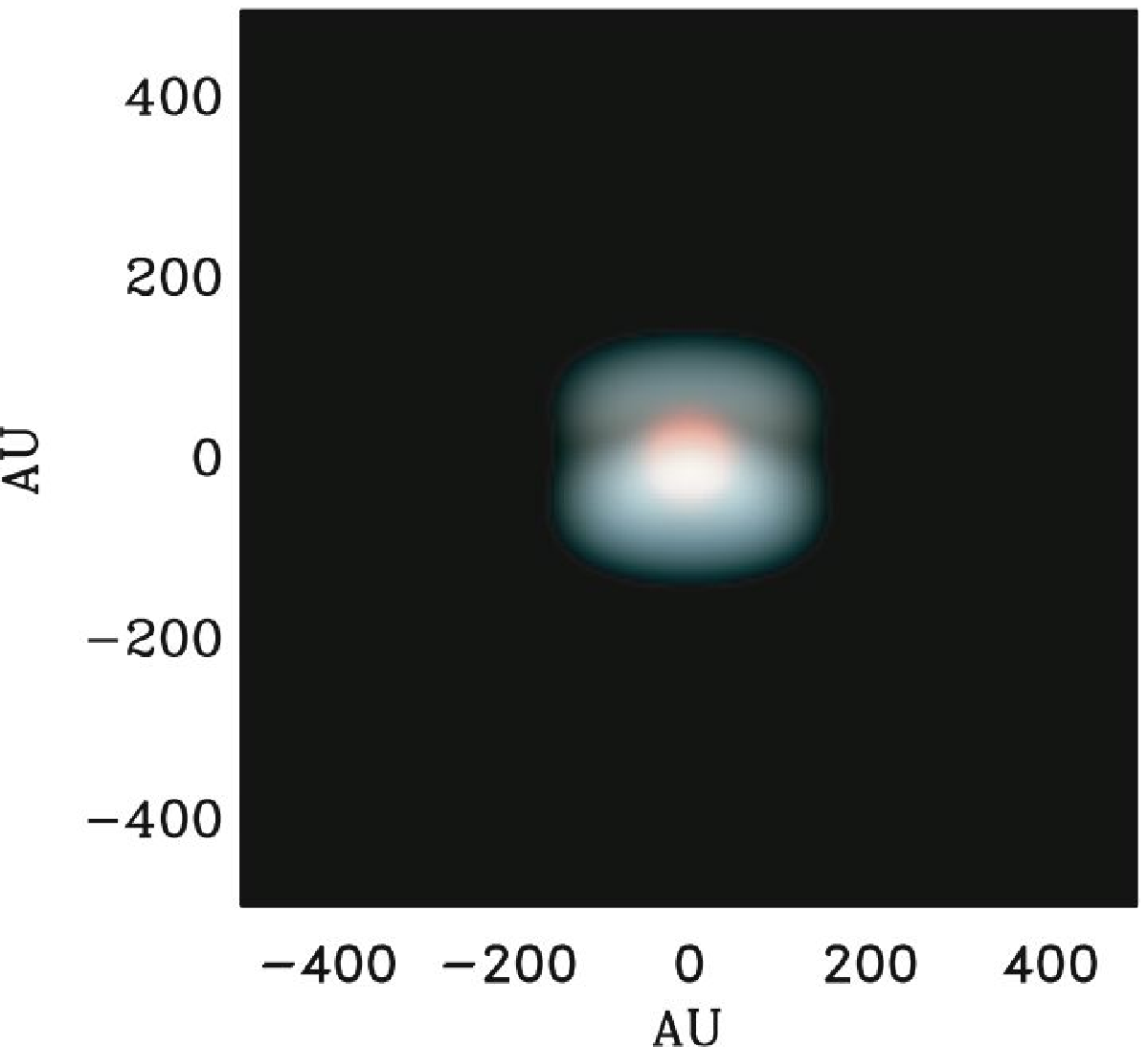}
  \epsscale{0.35}
  \plotone{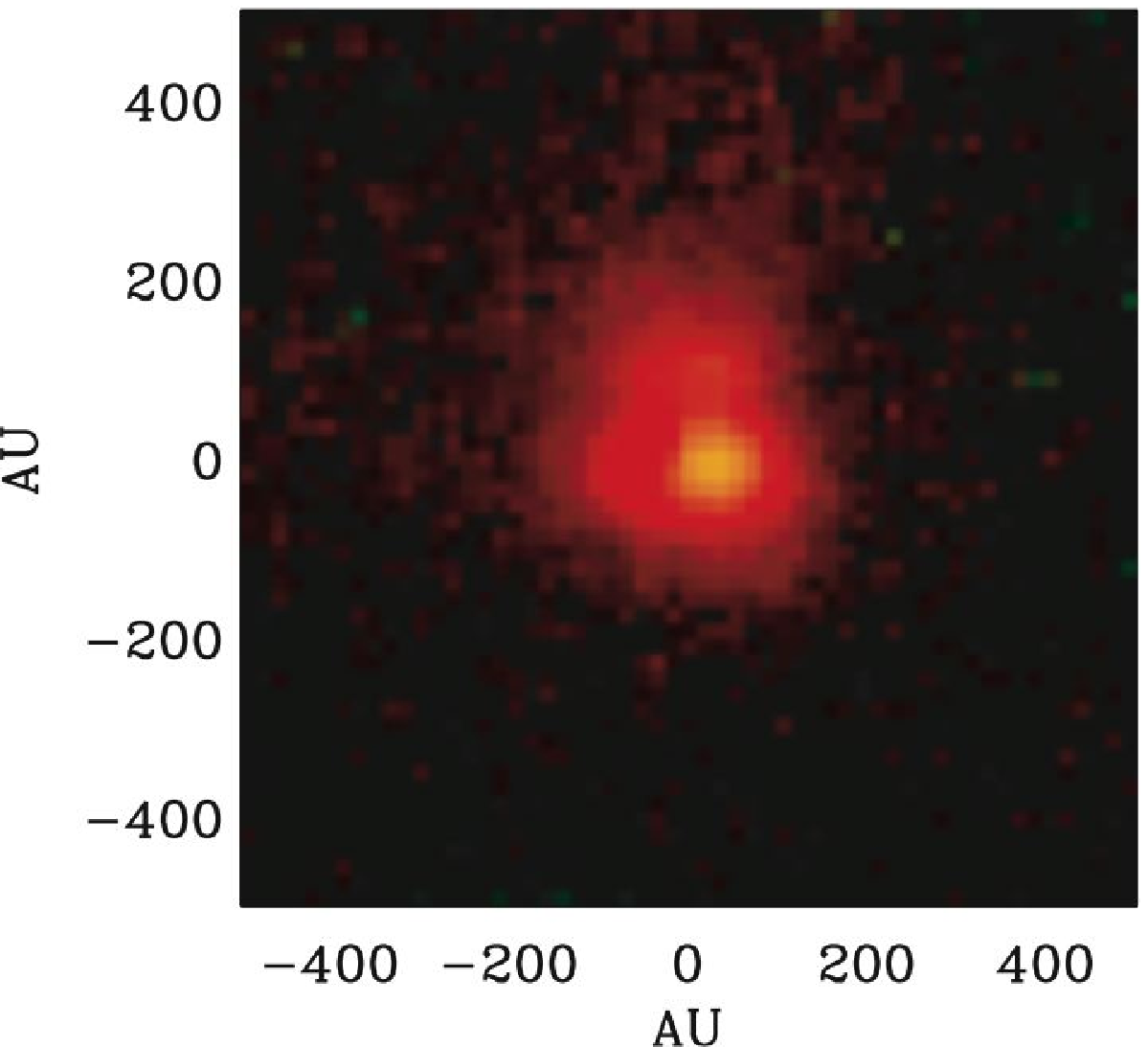}
  \epsscale{1.0}
  \caption{Observed and modeled composite $J$ (blue), $H$ (green), $K_s$ (red) images of CRBR 2422.8-3423. The color scale is logarithmic. The left side show the images of the disk
  as they appear including the foreground material, while the right side shows the images with the foreground material removed. Note that the central star and inner rim are seen as a reddened (by the disk material) point source in the center of the disk when the foreground is removed. The images are flux calibrated such that the observed and modeled colors can be directly compared. {\it Lower panel}: VLT-ISAAC color composite. {\it Middle panels}: $JHK_s$ model image corresponding to the SED in Fig. \ref{SED_MODEL}. The image has been convolved with a gaussian of $FWHM=0\farcs4$ to match the observed seeing. {\it Upper panels}: As above, but without convolving with the instrumental point spread function. }
  \label{JHK_MODEL}
\end{figure}
\clearpage

\begin{figure}
\plottwo{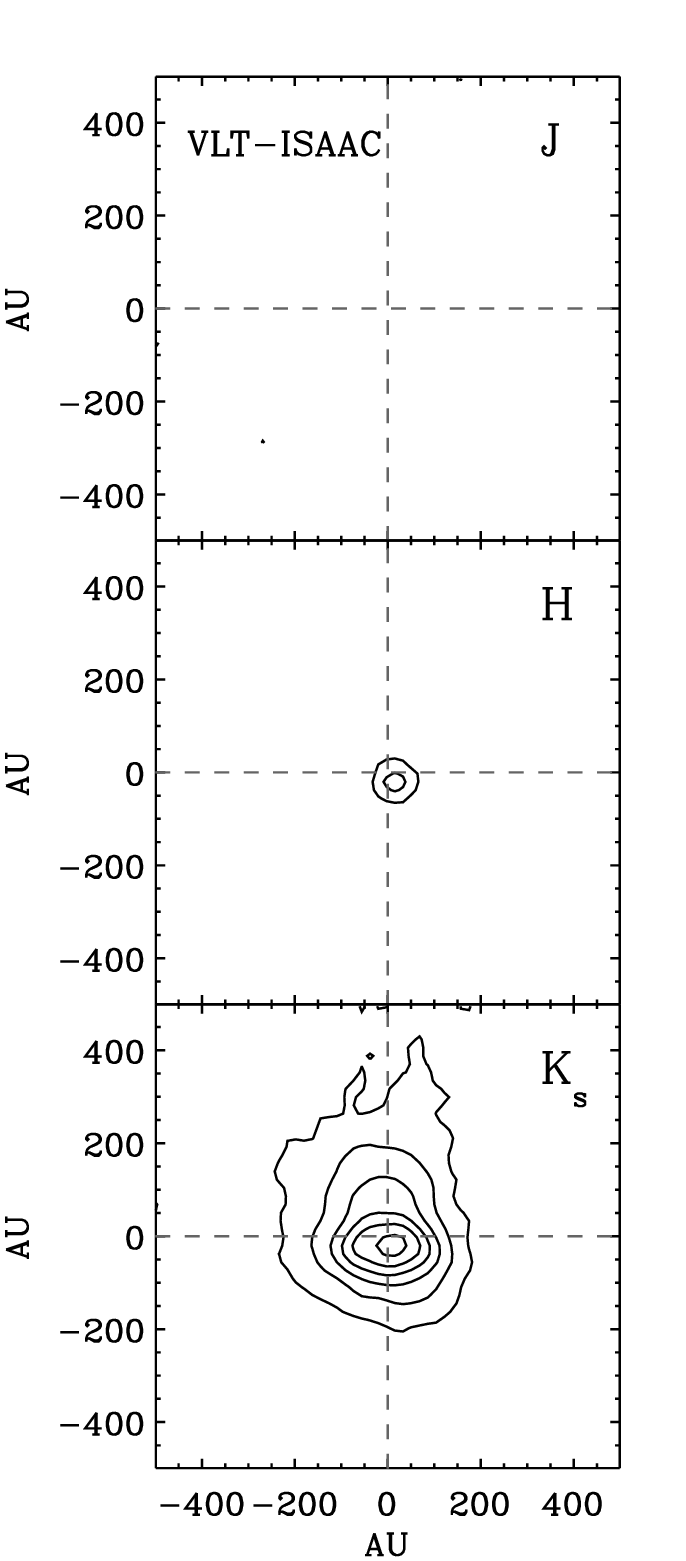}{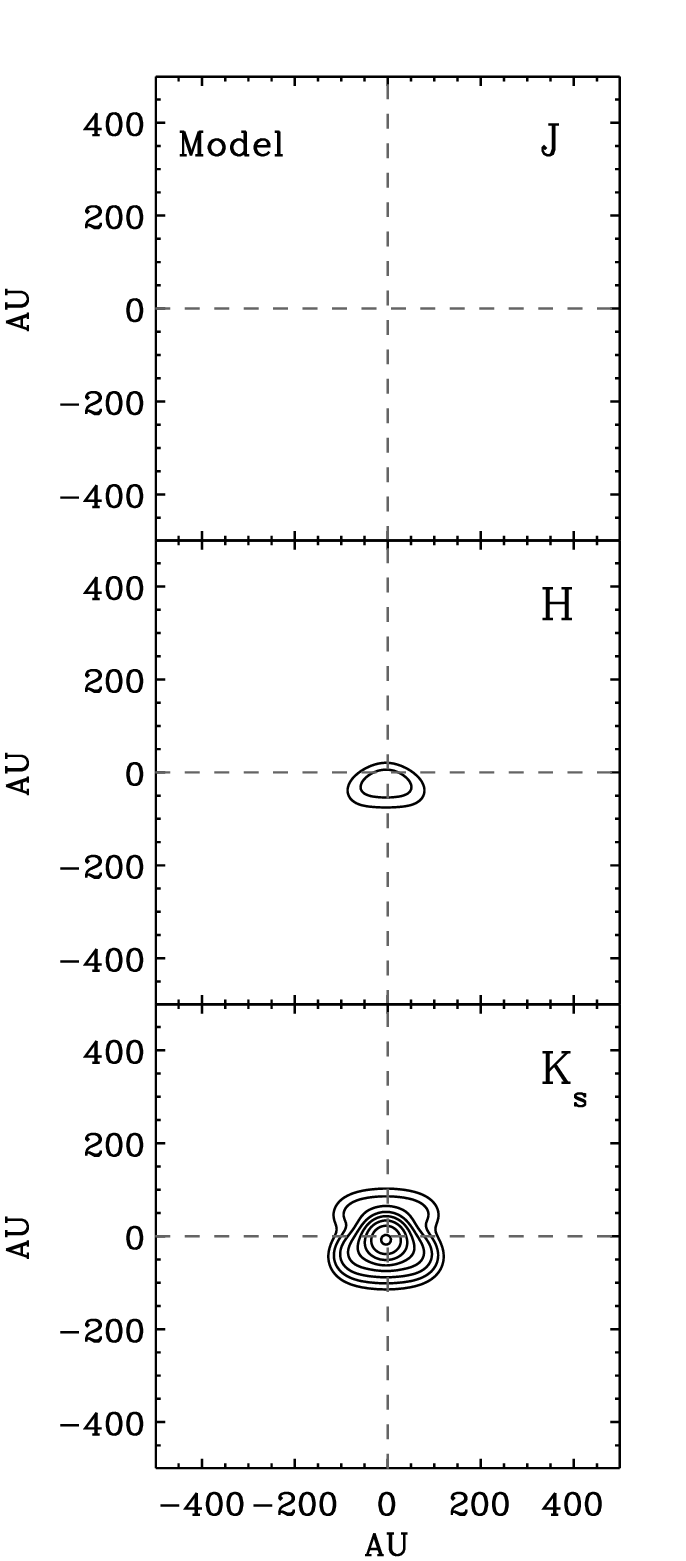}
\caption{{\it Left panels:} Contour plots of the VLT-ISAAC $JHK_s$ images of CRBR 2422.8-3423. {\it Right panels:} Contour
plots of the $JHK_s$ model images for the parameters given in Table \ref{parameters}. Note that the central star is visible in the model
images. The lowest contour is at $\rm 1.6\times 10^{-17}$\,erg\,s$^{-1}$\,cm$^{-2}$\,Hz$^{-1}$\,sterad$^{-1}$ and
subsequent contours increase in steps of 0.4 dex.}
\label{contours1}
\end{figure}

\begin{figure}
\plottwo{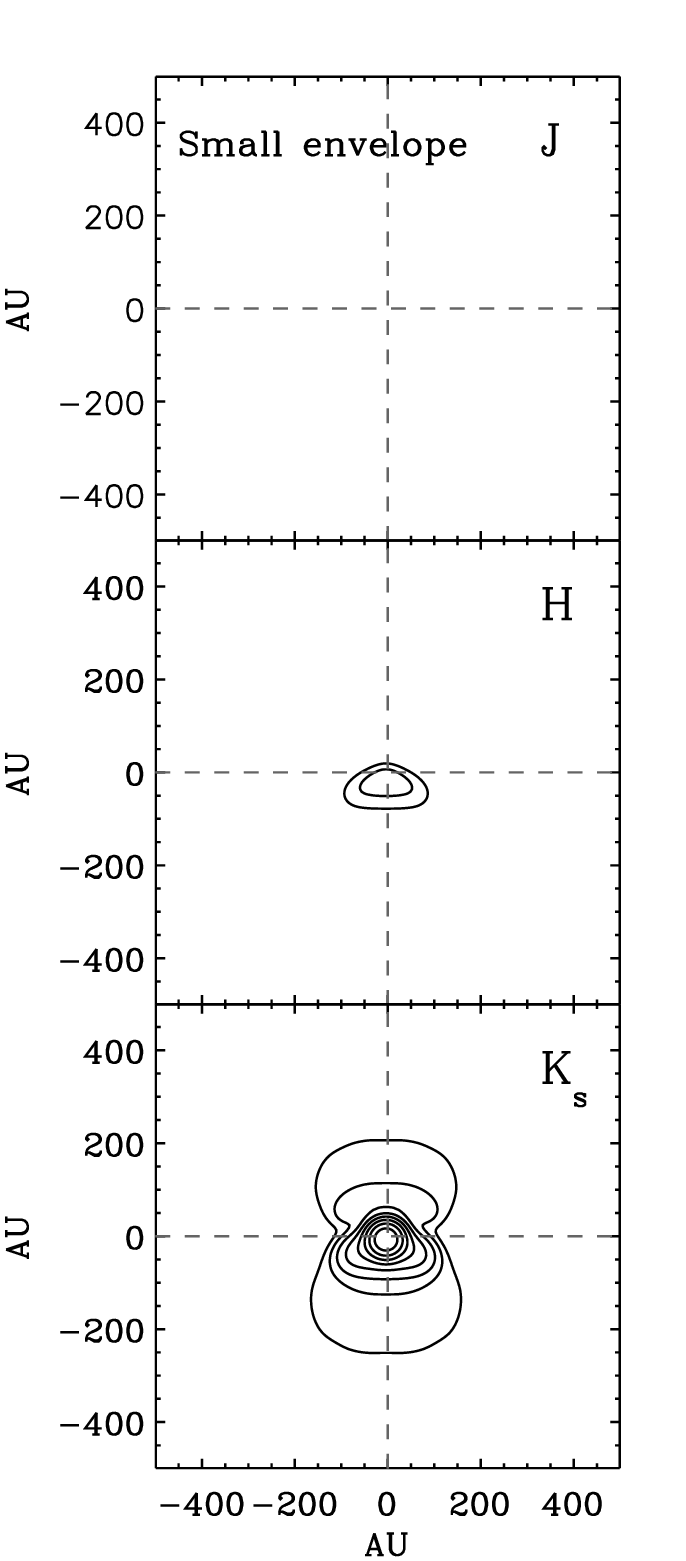}{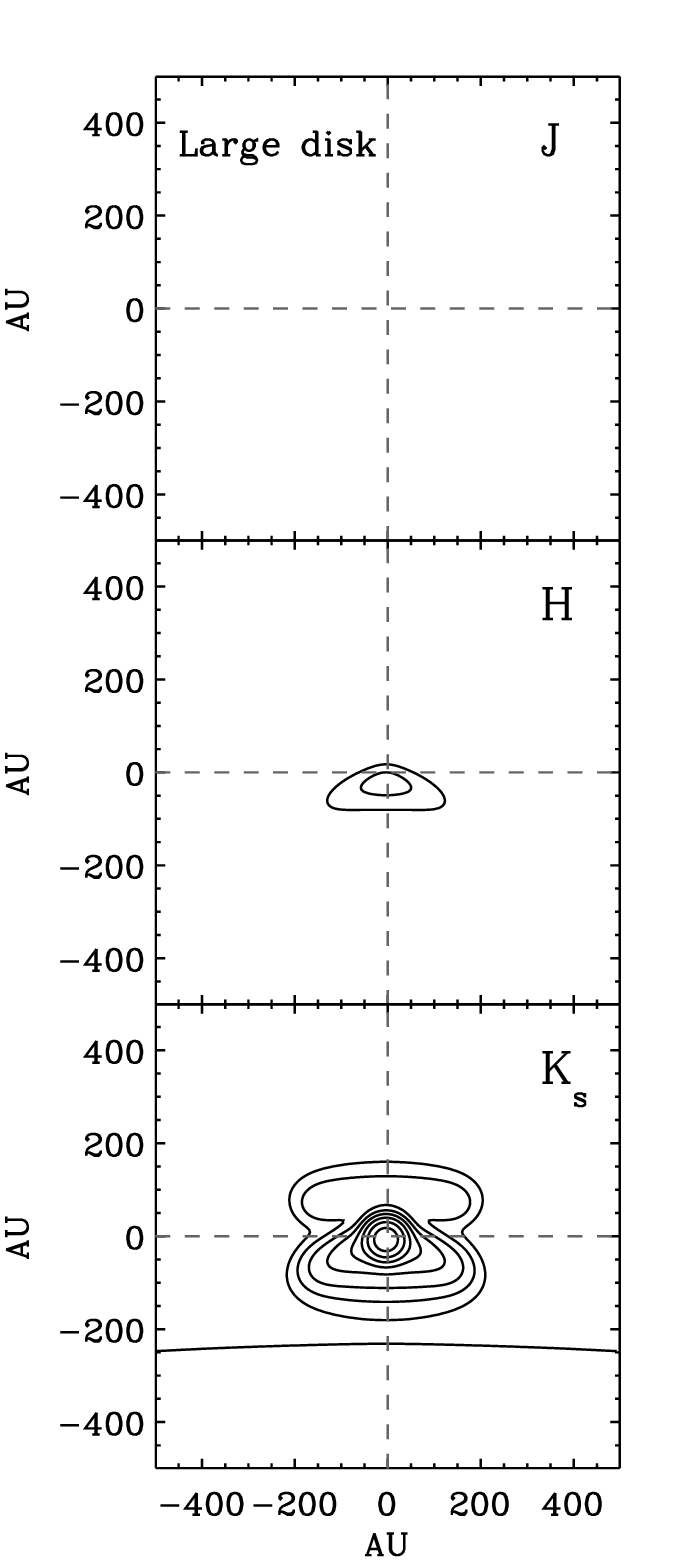}
\caption{{\it Left panels:} Contour plots of the $JHK_s$ model images for the `standard' model, but with a small power-law envelope added
to account for the excess nebulosity seen in the ISAAC imaging. {\it Right panels:} Contour
plots of the $JHK_s$ model images for a disk with $R_{\rm disk}=180\,$AU. The lowest contour is at $\rm 1.6\times 10^{-17}$\,erg\,s$^{-1}$\,cm$^{-2}$\,Hz$^{-1}$\,sterad$^{-1}$ and
subsequent contours increase in steps of 0.4 dex. }
\label{contours2}
\end{figure}

\begin{figure}
\plotone{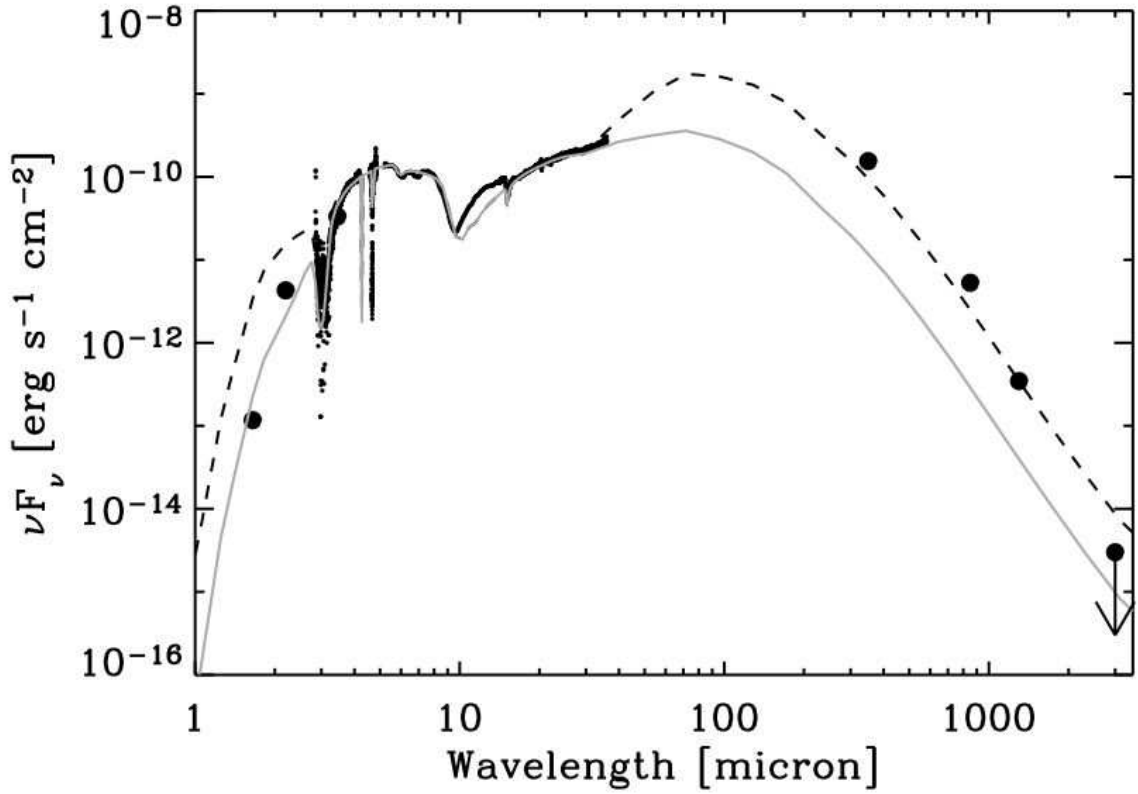}
  \caption{Best fitting SED model of CRBR 2422.8-3423. The solid gray curve shows the model flux through an
  aperture of 10\arcsec, while the dashed line shows the flux through an aperture of 30\arcsec. The 
  350\,$\mu$m, 850\,$\mu$m and 1300\,$\mu$m fluxes are integrated though a $30\arcsec$ aperture centered
  on the near-infrared position of CRBR 2422.8-3423. The 1300\,$\mu$m flux is taken from \cite{Motte98}.}
  \label{SED_MODEL}
\end{figure}

\begin{figure}
\plotone{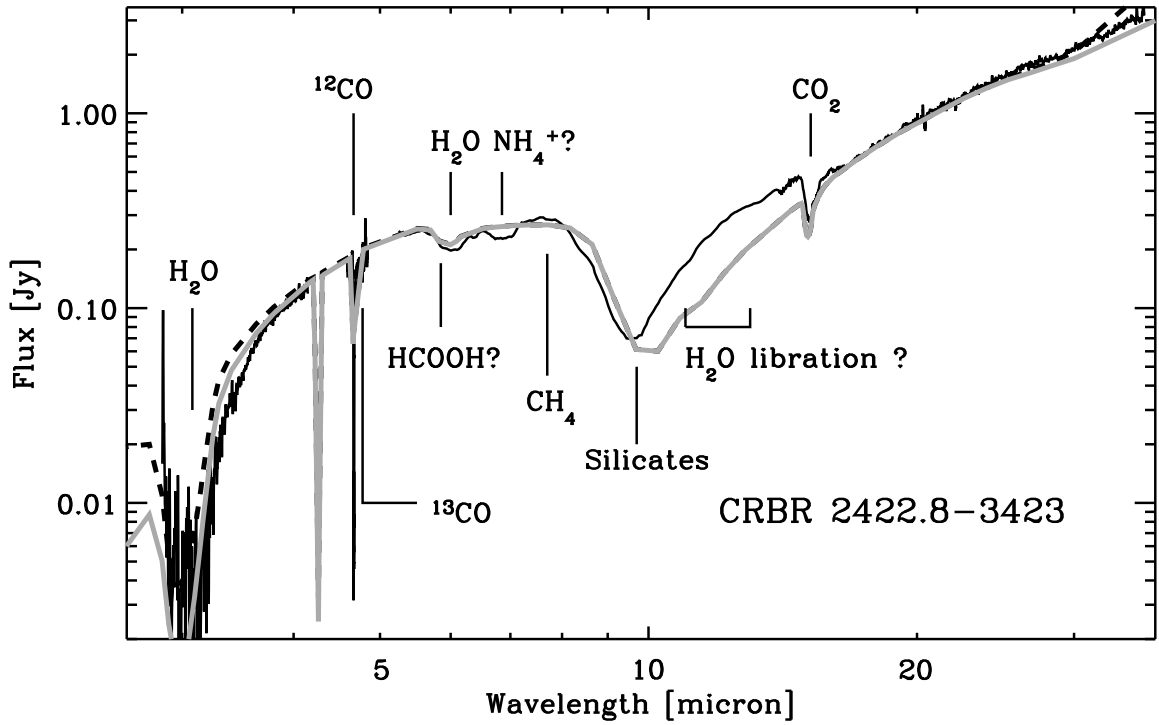}
  \caption{A close-up comparison of the best-fitting model in observing apertures of $10\arcsec$ (grey curve) and $30\arcsec$ (dashed curve) with the observed infrared spectra (black curve). 
  Note the mismatch on the red side of the 9.7\,$\mu$m silicate feature due to the water libration
  mode in the model.}
  \label{SED_MODEL_ZOOM}
\end{figure}

The `best fit' density structure shown in Fig. \ref{diskstruct} corresponds to a set of parameters which reproduce the observed constraints. Fig. \ref{JHK_MODEL} compares the synthetic $JHK_s$ composite images of the model disk to the observed image. Contours of the ISAAC images compared to the `best fit' model are shown in Fig. \ref{contours1}, while
contours showing alternative model images discussed in \S\ref{alternative} are shown in Fig. \ref{contours2}. 
Figs. \ref{SED_MODEL} and \ref{SED_MODEL_ZOOM} compare the modelled and observed SED.  The best-fitting parameters are summarized in Table \ref{parameters}. The table also includes estimates of the uncertainties of each model
parameter due to degeneracies. 

The model can, in principle, accommodate an arbitrary number of dust opacities, which can be made to depend on the dust temperature. This is useful for simulating the evaporation of different ices at different temperatures. For instance, a layer of solid pure CO desorbs at 20\,K at densities of $\sim 10^{6}\,$cm$^{-3}$, while water ice desorbs at 90\,K \citep{Sandford93}.
However, the desorption temperatures are very sensitive to the structure of the ice. CO may be trapped in
water ice and desorb at much higher temperatures than pure CO \citep{Collings03}, for example. 
Here we make a very simple model, assuming as a starting point that CO desorbs at 20\,K, since it is known to be largely pure, while all other ice species desorb at 90\,K.

The presence of several temperature-dependent opacities requires that the model is
iterated to obtain a self-consistent temperature structure. This was done by first calculating
a model with ice-covered grains throughout the disk. The output temperature structure was then 
used to determine where the ices evaporate and a new model was calculated with opacities uniquely defined at the relevant locations in the disk. 
This was repeated until the temperature became stable within 5\%. It was found that convergence is usually reached in the first few iterations. 

The parameters were varied manually to obtain the best fit because computing time prohibits the calculation
of a full grid. Any degeneracy in the best-fitting model parameters is therefore difficult to assess in detail. However, 
some parameters are clearly independent. As mentioned in \S\ref{DiskStruct}, the size of the flared disk
and the disk opening angle are constrained by the near-infrared images, as is the inclination angle. It is important to note that the disk may have a non-flared, self-shadowed component at large radii, which would not be seen in the scattered
light in the near-infrared. Thus, the near-infrared images constrain the size of the flared disk, but provide only a strict lower limit to the size of the entire disk if a hypothetical non-flared outer part is included. However, the current data cannot constrain a self-shadowed component. In particular, a flat outer disk will not contribute to the observed ice bands since the line of sight to the infrared source does not pass through such a flat disk component. Therefore only the flared part of the disk is modelled.  While not unambiguously determined, the total mass of the disk is constrained by the upper limit observed at 3\,mm in the OVRO interferometer to be $\lesssim 0.005\,\rm M_{\odot}$, using a dust mass opacity of $\kappa(3{\,\rm mm})=0.08\,\rm cm^{2}\,g^{-1}$. This upper limit also weakly constrains the possible mass of an outer self-shadowed disk component. 
\clearpage

\clearpage

A more realistic grain size distribution may significantly affect the mass estimate of the disk
\citep{Wood02}. However, the shape of the ice bands as well as the 9.7 $\mu$m silicate band indicate that the grains contributing to
the absorption bands are dominated by grain sizes of 0.1--1.0\,$\mu$m. This is due to the fact that scattering on 
grains with $2\pi a/\lambda_{\rm ice}\gg 1$, where $a$ is the size of the grain and $\lambda_{\rm ice}$ is the wavelength of the ice band in question, will enhance the contribution to the band from the real part of the refractive index, thereby creating excess absorption on the red side of each band. This is typically seen for the 3.08\,$\mu$m water ice band \citep[see e.g.][]{Leger83, Smith89, Dartois01}. For CRBR 2422.8-3423, the
red `scattering wings' of the ice bands are weak at wavelengths longer than 4\,$\mu$m, indicating that the contribution from grains with sizes $\gg 1\,\mu$m
is small. However, this may also simply indicate that the ice absorption is dominated by foreground material.

The most important indirect constraint on the disk structure is the high level of near to mid-infrared emission. Given the inclination angle, a disk mass of 0.01\,$\rm M_{\odot}$ or larger produces too much extinction along the line of sight to the inner disk to match the Spitzer spectrum. Even with the lower mass used in the presented model, it is found that the presence of an inner rim is essential to reproduce the shape and high flux level of the mid-infrared spectrum. 
Alternatively, unrealistically lowering the effective temperature of the central star to 2000\,K produces a similar effect. An important point in this context concerns the visibility of the central star in the near-infrared images. At an inclination of $\sim 90\degr \times (1-H_{\rm disk}/R_{\rm disk})$, the inner disk and star become visible to the observer. This usually results in the star outshining the disk by orders of magnitude. This is clearly not the case for CRBR 2422.8-3423 and is the
reason the source was first detected as an edge-on disk. However, the blocking of the star by the
flared disk also tends to strongly suppress the mid-infrared continuum. Therefore, in order to fit both
the images as well as the Spitzer spectrum, only a very narrow range of surface density profiles, and consequently disk masses, are possible, given an opacity. The puffed-up inner rim is vital for blocking 
the light from the central star sufficiently without suppressing the mid-infrared continuum, although even in our best fit model some of the stellar light can still be seen directly in excess of what is observed in the $K$-band;
in the model the star is about 10 times too bright at 2.2\,$\mu$m. However, in the $J$- and $H$-bands the model
star is much fainter than the disk nebulosity in accordance with the observations. Furthermore, the bright mid-infrared continuum clearly indicates that the central parts of the disk dominate over scattered light at wavelengths longer than $\sim$3\,$\mu$m. Thus, the problem is confined to the $K$-band image. This is most clear in Fig. \ref{JHK_MODEL} where the foreground component has been removed. Here a reddened (by the disk material) point source is visible in the center of the disk. The fact that the central star in the model cannot be completely suppressed at 2\,$\mu$m while preserving the high mid-infrared continuum is a significant short-coming of the model. Possible explanations include a different near-infrared opacity law or a different geometry of the inner disk, in particular the puffed-up inner rim. The brightness ratio of 1:11 \citep{Brandner00} of the two scattering lobes of the disk was therefore fitted by comparing the modeled and observed $K$-band images along a line offset by 50\,AU along the major axis of the disk to avoid the light of the central star. The `best-fitting' model brightness ratio is 1:12 along this offset line.

The column density of foreground material toward the disk is well constrained by the near-infrared color to $2.2\times 10^{22}\,$H$_2$\,cm$^{-2}$, assuming
a gas-to-dust mass ratio of 100, corresponding to an extinction in the $J$-band of 8\,mag.
If this material is placed close to CRBR 2422.8-3423, for instance in the form of a remnant envelope, a bright reflection nebulosity is created. Since no such extended reflection nebulosity is seen in the $K$-band imaging, the foreground material must be located at a distance of at least 1000\,AU from the source. This is consistent with the line of sight passing through the Oph-F core seen in the submillimeter imaging, rather than an envelope surrounding CRBR 2422.8-3423. Alternatively, the source may already have excavated a significant cavity in its envelope. A small amount of excess nebulosity compared to the `best fit' model within 300\,AU is seen in Fig. \ref{contours1}. The possible presence of a small tenuous envelope of CRBR 2422.8-3423 is discussed in \S\ref{alternative}. The lack of bright, extended scattering nebulosity in the near-infrared imaging puts an upper limit of a few$\times\,10^4$ for the density within 1000\,AU of CRBR 2422.8-3423. The density of the material producing the near-infrared extinction was constrained by fitting the submillimeter emission within an aperture of {30\arcsec} given the column density calculated using the near-infrared colors. The arbitrary method of fitting the foreground material using a spherically symmetric envelope introduces an uncertainty of a factor of 2 in the density. This is because the source may be located entirely behind the material producing the submillimeter emission and not in the middle, such as is assumed in the model.
 
\clearpage
\begin{table*}[h!]
\footnotesize
\centering
\begin{flushleft}
\caption{Best-fitting model parameters}
\begin{tabular}{lll}
\hline
\hline
&Actual model value&Estimated range (see text)\\
\hline
$T_{*}$&4500\,K&3500--5500\,K\\
$R_{*}$&$2.0\,R_{\odot}$&1.5--2.5\,$R_{\odot}$\\
$M_{*}$&0.8\,$M_{\odot}$&0.2--2.0\,$M_{\odot}$\\
$L_{*}$&1.4\,$L_{\odot}$&1.3-1.5\,$L_{\odot}$\\
$M_{\rm disk}$&$0.0015\,M_{\odot}$&0.0005--0.005\,$M_{\odot}$\\
$R_{\rm disk}$&90\,AU& 80--100\,AU (Only flared disk)\\
$H_{\rm disk}/R_{\rm disk}$&0.18&0.16--0.20\\
$H_{\rm rim}/R_{\rm rim}$&0.12&0.10--0.14\\
$R_{\rm in, env}$&1000\,AU&$\gtrsim 1000$\,AU\\
$R_{\rm out, env}$&6700\,AU&$\lesssim 7000$\,AU\\
$\rho_{\rm env}$&$2.6\times 10^5\,$cm$^{-3}$&$1.3-3.0\times 10^5\,$cm$^{-3}$\\
Incl. angle&69.2\degr&68--72\degr\\
H$_2$O ice abundance (env.+disk) & $9\times 10^{-5}$ (rel. to H$_2$)&$8-10\times 10^{-5}$\\
CO ice abundance (envelope) & $1.3\times 10^{-4}$ (rel. to H$_2$)&Assuming no CO ice in the disk\\
\hline
\end{tabular}
\label{parameters}
\end{flushleft}
\end{table*}
\clearpage

\clearpage

\subsection{Alternative models}
\label{alternative}
\begin{figure}
\plotone{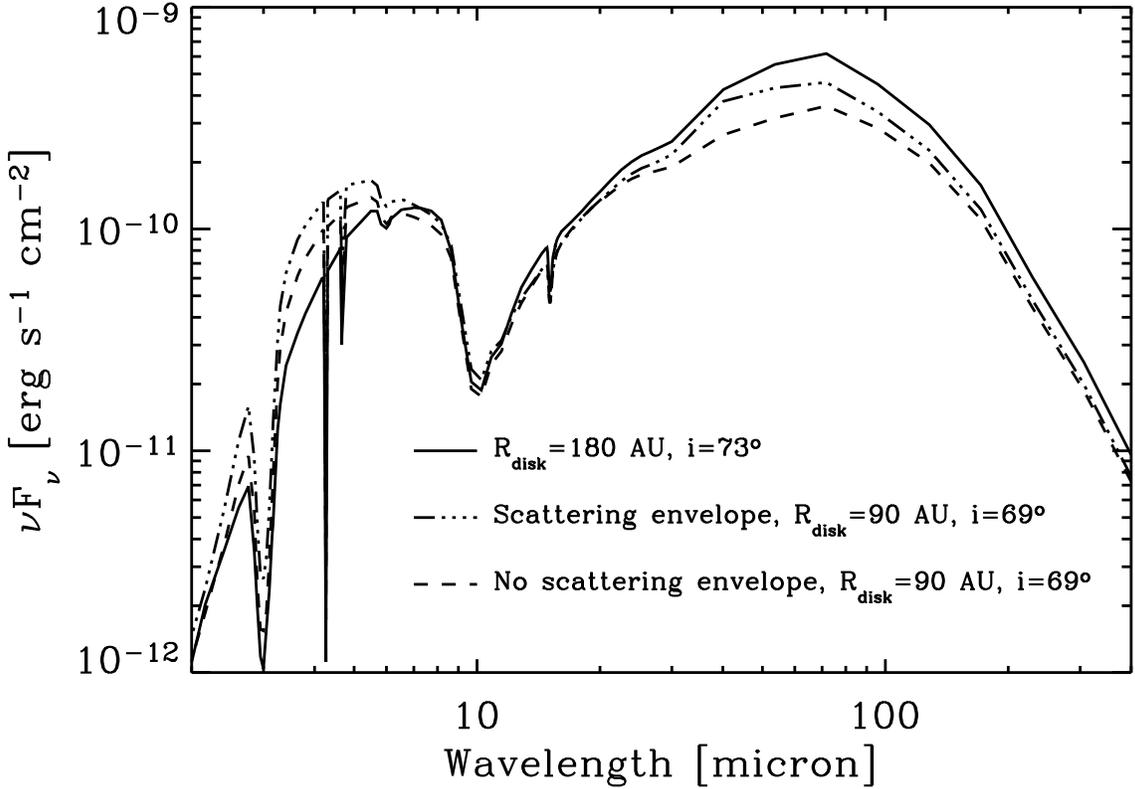}
  \caption{Comparison of the SEDs for different models of CRBR 2422.8-3423. The dashed curve shows the `standard' model
  fit presented in Fig. \ref{SED_MODEL}. The dash-dotted curve shows the SED if a small power-law envelope has been added to
  account for the excess nebulosity seen in the ISAAC images of CRBR 2422.8-3423. The solid curve shows the SED of a 
  model with twice the disk radius of the `standard' model (180\,AU). This model has an inclination of $73\degr$ and an increased stellar luminosity
  of 2.9\,$L_{\odot}$ to fit the observed SED. }
  \label{SED_MODCOMP}
\end{figure}

In this section, we explore the sensitivity of the model to simultaneous SED and imaging constraints. Specifically, we explore if the temperature of the absorbing material in the disk can be lower than in the `best-fitting' model in order to facilitate the formation of CO ice. In this context, an important question
is whether the size of the flared disk is well-constrained by the ISAAC image, since a larger disk may contain colder dust. To test this, a model was calculated by assuming $R_{\rm disk}=180\,$AU,
but fixing the remaining structural parameters to those in the `standard' model given in Table \ref{parameters}. A best fit to the SED was then obtained by varying the luminosity of the central star and the inclination of the system. Because the mass of the disk is kept constant the 
inclination needs to be higher to produce the same extinction toward the star. At the same time, the fraction of warm dust in the disk decreases, requiring the luminosity to be increased to fit the mid-infrared emission. This uncertainty on the luminosity is primarily due
to the lack of far-infrared photometry. The resulting image is seen in Fig. \ref{contours2} and the SED is compared
to the `standard' model in Fig \ref{SED_MODCOMP}. It is seen that the SED of the large disk can still provide a good
fit to the data by varying only the inclination and luminosity. However, the $K_s$ image of the large disk clearly becomes larger than 
the observed image. We conclude that the size of the flared disk is constrained by the near-infrared imaging. Modeling of an outer
non-flared (self-shadowed) disk is postponed to a later paper. 

The ISAAC $K_s$ image also clearly shows excess scattering emission close to the central disk (Fig. \ref{contours1}). This emission is not included in the `standard'
model. A small envelope may affect the modelled flaring properties of the disk by raising the scattering surface \citep{Stapelfeldt03}. 
Here, we test the effect on the model image and SED of a small envelope producing excess emission. A small power-law envelope is added
at radii larger than $R_{\rm disk}=90\,$AU to fit the ISAAC images. An envelope with a density power-law exponent of -1 and
a gas density at $R_{\rm disk}$ of $10^{4}$\,cm$^{-3}$ provides a reasonable fit to the excess emission in the ISAAC $K_s$ image. Note, however, that the shape of the density profile is not strongly constrained. The influence of the small scattering envelope on the SED is seen in Fig.
\ref{SED_MODCOMP} to be minimal. Additionally, the column density of such a remnant envelope is so small that it has no influence on the ice absorption bands
in the model.

\clearpage
\clearpage

Finally, we explore the possibility that the disk may be more inclined than $\sim70\degr$. There are two ways of increasing the model inclination while preserving the spectral characteristics of the source, in particular the high mid-infrared continuum: the first is to decrease the disk opening angle and the second is to decrease the mass of the disk. If the disk opening angle is decreased by a factor of 2 to 0.09, the inclination can be increased to $\sim 76\degr$. However, since the disk intercepts less
stellar irradiation, the luminosity of the central star has to be increased significantly, causing absorbing disk material to increase its temperature to the levels of
the less inclined disk. Additionally, the disk becomes too flat to fit the near-infrared images. If the mass of the disk is decreased to facilitate are higher inclination of 75\degr, the temperature of the disk mid-plane increases to well above 30\,K.

\section{Where are the ices toward CRBR 2422.8-3423 located?}
\label{icesindisk}

\subsection{Ices in the model disk}
\begin{figure}
\plotone{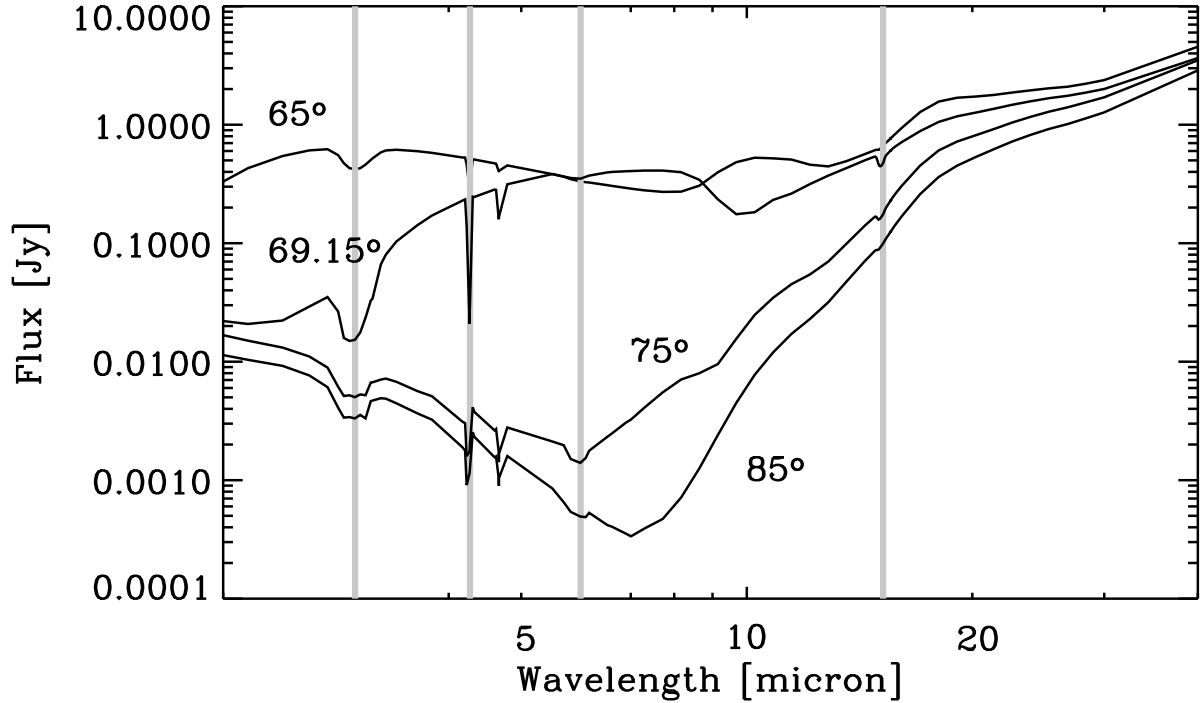}
  \caption{Models showing the effect of inclination on the ice bands. The models are those that best-fit CRBR 2422.8-3423, only with the foreground cloud removed. The inclinations shown are from top to bottom: $65\degr$, $69.15\degr$ (`standard' model value - see Table \ref{parameters}), $75\degr$ and $85\degr$.  Note the significant changes in line ratios, especially between the 3.08\,$\mu$m and the 6.0\,$\mu$m water ice bands. The vertical lines
  indicate the position of the 3.08\,$\mu$m water stretching mode, the 4.27\,$\mu$m CO$_2$ stretching mode, the 6.0\,$\mu$m water bending mode and the 15.2\,$\mu$m CO$_2$ bending mode.}
  \label{SED_MODEL_NOFRONT_incl}
\end{figure}

The radiative transfer model of CRBR 2422.8-3423 allows the location
of the ices to be studied in greater detail. In principle, the model cannot directly distinguish between ice located in the disk and ice located in foreground material, but significant conclusions can be reached by considering the range of realistic ice abundances. Furthermore, differences in dust temperatures in the disk and foreground place constraints on the location of the ices.  Fig. \ref{SED_MODEL_NOFRONT_incl} shows the SED for the disk with the foreground material removed. It is seen that significant fractions of the ice absorption lines originate in the disk material if the assumption of a constant water ice abundance of $9\times 10^{-5}$ in the disk and foreground material is correct. Quantitatively, 
the model disk contributes $45\%$ of the observed water ice band for a constant ice abundance.

\begin{figure}
\plotone{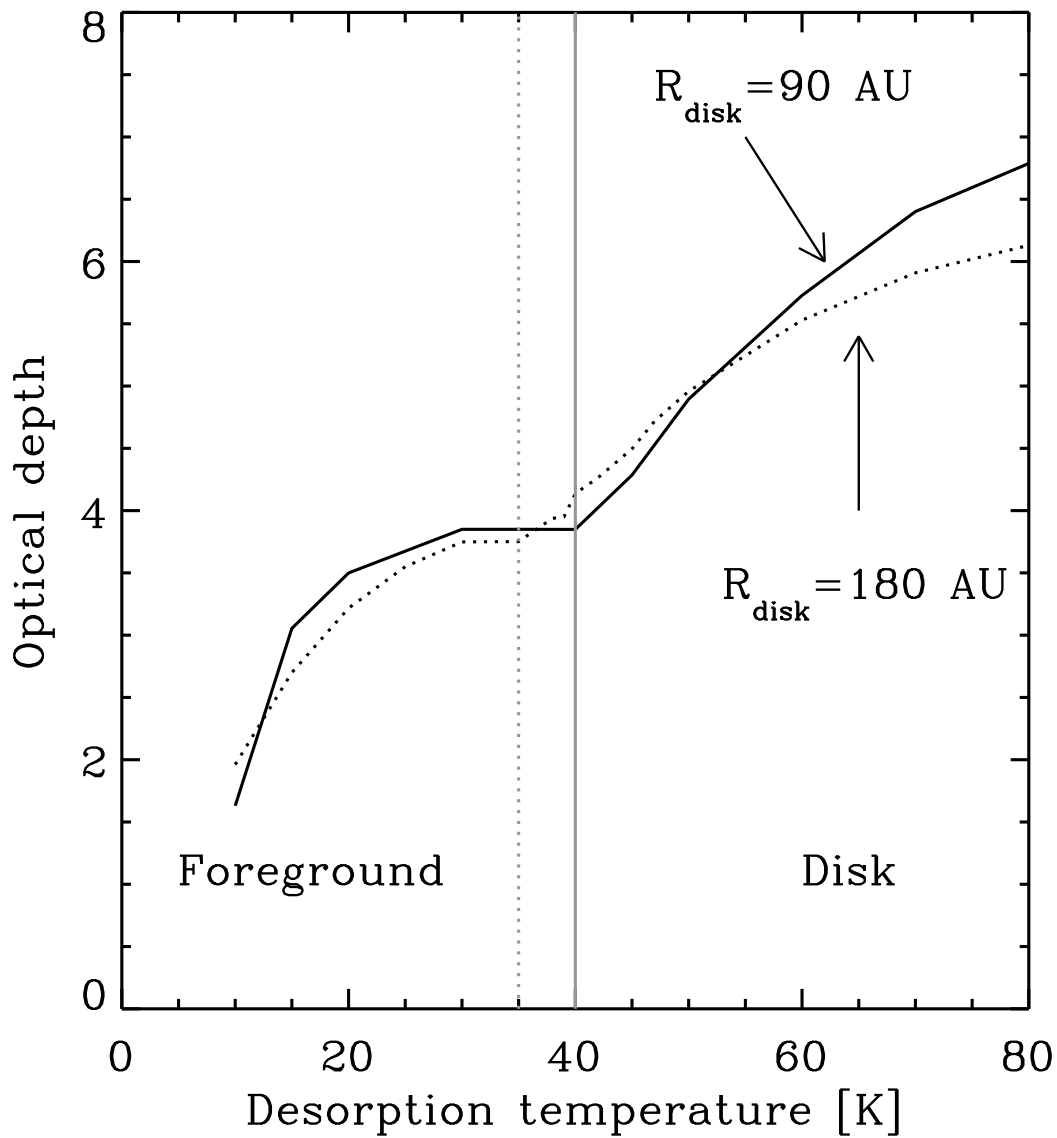}
  \caption{Optical depth of the 4.67\,$\mu$m band of CO for the CRBR 2422.8-3423 model as a function of (the assumed) desorption temperature. The abundance is arbitrarily fixed at $9\times 10^{-5}$ whenever the temperature is below the desorption temperature. Both the `standard' model (solid curve) of Table \ref{parameters} and the model with $R_{\rm disk}=180\,$AU (dotted curve) are shown. The vertical solid and dotted lines indicate the separation between contributions from the foreground material and from the disk material for the 90\,AU and 180\,AU disks, respectively. }
  \label{SED_MODEL_CO}
\end{figure}

From Fig. \ref{diskstruct} it is seen that the temperature in the model disk nowhere drops below 27\,K.
This effectively prevents any pure CO ice from being present in the model disk, since this evaporates at
$\sim 20$\,K at the densities present in the disk \citep{Sandford93}. A significant change of the disk density structure, such as 1--2 orders of magnitude denser material in the mid-plane or a significantly larger outer disk radius, is required in order to
have pure CO ice be stable in the disk. Even if CO ice is present in a dense mid-plane, it is unlikely to contribute to the observed CO ice band. This 
question can be explored by examining the temperature of the dust that can be probed at a wavelength of 4.67\,$\mu$m. Fig. \ref{SED_MODEL_CO} illustrates the point for the best-fitting model disk. It shows the model optical depth of the 4.67\,$\mu$m CO band as a function of desorption temperature, assuming a constant CO ice abundance for temperatures lower than the desorption temperature. The relation has been calculated by directly including a component of pure CO inclusions in the dust model for temperatures lower than the desorption temperature. Naturally, if a higher desorption temperature is assumed, a deeper CO ice band appears in the spectrum. Therefore, the contribution of ice at different temperatures to an ice band centered on 4.67\,$\mu$m can be read off the relation in the figure. The foreground material is seen as a sharp rise to
20\,K followed by a plateau, indicating that little absorbing material is present at temperatures between 30 and 40\,K.
The contribution from material in the disk takes over only at temperatures above 40\,K. This clearly shows that the model
disk can contain no pure CO ice that will show up in the spectrum; all ice in the disk observed in the wavelength region around 5\,$\mu$m must have temperatures between 40 and 90\,K. This is consistent with the excitation temperature of the observed CO rovibrational lines of 40--60\,K as determined by \cite{Thi02}. Inspection of Fig. \ref{diskstruct} shows that the disk material, which contains ices contributing to the observed absorption bands, has densities of $10^6-10^7\,$cm$^{-3}$. The disk must be considerably larger in order to contribute to the observed pure CO ice
band. This is illustrated by plotting the CO ice band depth as a function of assumed desorption temperature of the large disk with $R_{\rm disk}=180$\,AU in Fig. \ref{SED_MODEL_CO}. It is seen that the large disk shifts the absorbing material toward lower temperatures as expected, but
only by $\sim 5\,$K, such that the coldest dust in the disk contributing to the ice band is 35\,K for the large disk. 

If the CO is embedded in other less volatile ice species, such as water ice, the evaporation temperature of CO can be raised to 70\,K. However, the CO and CO$_2$ ice profiles indicate that at most 20\% of the CO is mixed with CO$_2$ and another 20\% with water. The remaining
CO ice must reside in the cold foreground dust. 
The conclusion that none of the pure CO ice observed in the spectrum is present
in the disk implies that the abundance of pure CO ice in the foreground must be very high, $1.1\times 10^{-4}$, making CO the most abundant ice species in the foreground material. If the CO embedded in water ice is present in the observed average abundance of $1.7\times 10^{-5}$ (see Table \ref{abundances}) throughout the system, the CO ice abundance in the foreground may increase further to $1.27\times 10^{-4}$. In order to determine the total depletion of gas-phase CO in the foreground, ice species formed by CO, such as CO$_2$ and CH$_3$OH, must also be included. Adding the observed average abundance of CO$_2$ of $2.9\times 10^{-5}$ and ignoring the relatively small contribution from CH$_3$OH, the total amount of CO depleted from the gas-phase is as much as $1.6\times 10^{-4}$. 
For a total CO abundance of $2-3\times 10^{-4}$ w.r.t. H$_2$, this leaves at most $1\times 10^{-4}$ for gas-phase CO, corresponding to a foreground column density of $2.2\times 10^{18}\,$cm$^{-2}$. This is roughly consistent with the total
CO column density (foreground + background) of $\sim 5\times 10^{18}\,$cm$^{-2}$ derived from
millimeter C$^{18}$O data \citep{Thi02}.
We conclude that the CO gas in the foreground cloud is depleted 
by a factor of 2--3.

\begin{figure}
\plotone{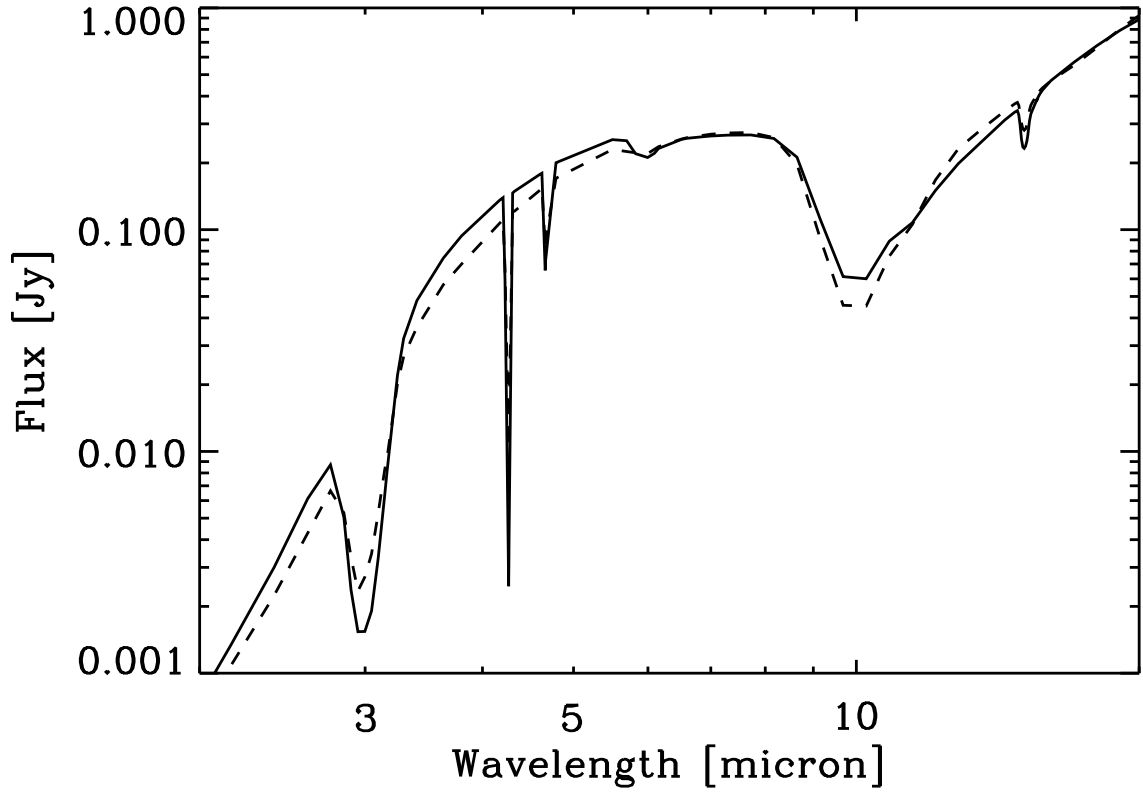}
  \caption{Comparison of a model assuming that the ice evaporation temperature
  is  $T_{\rm evap}=90$\,K (solid curve) with a model using $T_{\rm evap}=45\,$K (dashed curve). For CRBR 2422.8-3423, a significant amount of the ice has temperatures in excess of 45\,K.}
  \label{SED_MODEL_lowtemp}
\end{figure}

The assumption that all other ice species evaporate at 90\,K is not very realistic and was adopted 
for simplicity. In reality, the ices considered in the model evaporate at a range of temperatures between 20 and 100\,K, depending on the species and ice composition. 
Fig. \ref{SED_MODEL_lowtemp} shows the variation of the model ice features in the disk if an evaporation temperature of 45\,K is assumed instead. It is seen that the ice bands become $\sim 30\%$ weaker
for the lower evaporation temperature, corresponding to most of the disk ice. The implication is that all disk ice features should show significant signatures of heating above 40\,K. For CRBR 2422.8-3423, this effect may be difficult to observe due to the contribution from cold foreground ice. However, as discussed in Sec. \ref{5to8}, 
the red shoulder of the 6.85\,$\mu$m band indeed provides significant evidence for the presence of warm ices along the line of sight. It is important to note that most of the disk mass has temperatures below 40\,K, but that this material is simply
not probed by the ice bands due to the high optical depths in the mid-plane and the system inclination angle.  
This illustrates how the location of ices in circumstellar disks and probably in most low-mass protostars require a detailed axisymmetric model to be adequately constrained.

In summary, we find that up to 50\% of the observed band optical depths of ice species that remain in the solid state at temperatures above 40\,K, may be due to disk ice. This includes water, CO$_2$ ices and CO embedded in water ice. The observed pure CO ice is most likely located entirely in the Oph-F core.

\clearpage

\clearpage

\subsection{The missing water ice librational band}
In Fig. \ref{SED_MODEL}, it is evident that the red wing of the 9.7\,$\mu$m silicate band is not well-fitted by the model. The model red wing is caused by the librational (hindered rotation) band of water ice. This band is centered around 12-13\,$\mu$m, but the center and width are sensitive to ice composition \citep{Hagen83}. The band strength of the water libration band is almost three times higher than the CO$_2$ bending mode, which is located in the same wavelength region \citep{Gerakines95}. Combined with the higher column density of water ice, the libration band should be about 10 times stronger than the corresponding 15.2\,$\mu$m CO$_2$ band, although it will only contribute an optical depth of 0.3-0.5 at 12\,$\mu$m due to the large width of the band. Laboratory experiments find that the strength of the librational band is fairly constant under interstellar conditions \citep{Kitta83}. Careful radiative transfer models of other sources compared to, e.g., Spitzer-IRS spectroscopy may show if the problem is universal. Recent Spitzer-IRS spectra of the embedded low-mass star HH 46 IRS, toward which the ice bands are about twice as deep as toward CRBR 2422.8-3423, also do
not show a water ice libration band as strong as expected \citep{Boogert04}. The librational band is sensitive to the dust model, through the grain size distribution as well as the shape of the grains. It is possible that a more realistic dust model may provide a better fit the librational water ice band.

\section{Observing ices in other highly inclined disks}
\label{Others}

Using the model for the CRBR 2422.8-3423 disk, some general predictions can be made for 
observations of ices in other edge-on circumstellar disks. Clearly, the presence of cold foreground material complicates any interpretation of ice bands from a disk. However, since this problem often cannot be avoided, it is interesting to explore how the ice bands intrinsic to the disk may be distinguished from those produced by foreground material.
\clearpage
\begin{figure}
\epsscale{0.6}
\plotone{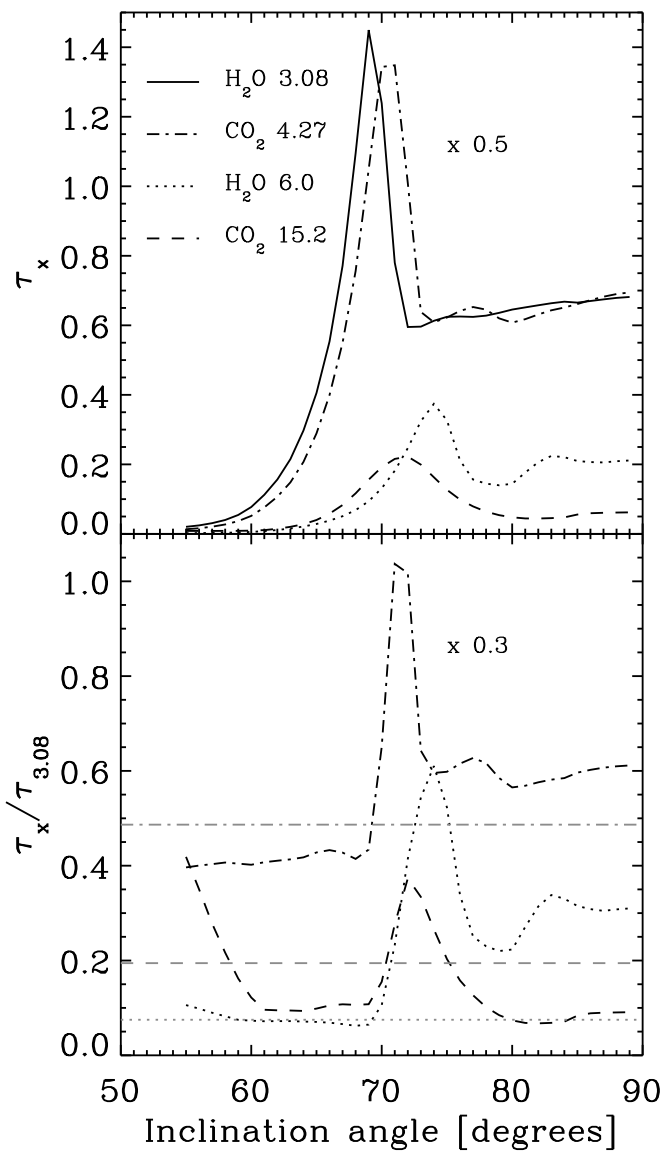}
  \caption{{\it Upper panel}: Optical depths of the H$_2$O and CO$_2$ ice bands as functions of disk
  inclination for the model of CRBR 2422.8-3423. {\it Lower panel}: Line ratios relative to the 3.08\,$\mu$m water ice band. The horizontal lines indicate the ratios of the extinction coefficients of the bands, i.e. the line ratio expected to be observed in a cold foreground cloud. No foreground material has been included to clarify the effects intrinsic to the disk. The curves for the CO$_2$ 4.27\,$\mu$m band
  have been scaled by a factor of 0.5 (upper panel) and 0.3 (lower panel) to fit in the figure.}
  \label{icefeatures_incl}
\end{figure}
\clearpage

An interesting question to explore is how the ice bands behave with varying inclination. For instance, are disks with an inclination angle close to $90\degr$ more favorable for observing ices than disks with lower inclination angles of $\sim$70\degr?
In Fig. \ref{SED_MODEL_NOFRONT_incl} the spectra of the best fitting model of CRBR 2422.8-3423 viewed at different inclination angles are shown. It is apparent that an inclination of $70\degr$ in fact produces the deepest ice bands. A disk similar to CRBR 2422.8-3423 viewed very close to edge-on shows only weak intrinsic ice features. This is an effect produced by the large optical depth through the mid-plane. Disks 1--2 orders of magnitude less massive than CRBR 2422.8-3423 will
have optical depths approaching unity along the mid-plane. Such an edge-on disk will have deep ice bands if ices are present in abundance.

Another observation of considerable interest is that the ratios between different ice bands of the same species are not preserved as a function of inclination.  To illustrate this quantitatively, the model optical depths of H$_2$O and CO$_2$ bands are plotted in Fig. \ref{icefeatures_incl}  as functions of the
disk inclination. Common for all ice bands is that the optical depth peaks strongly for inclinations between 65 and 75\degr. For most ice bands a constant plateau is reached at higher inclination angles
after the main peak. The exception here is the 15.2\,$\mu$m CO$_2$ bending mode, which
drops to a very small absorption depth at the highest inclination. Finally, the different bands are seen to
peak at somewhat different inclination angles.

The model optical depths are also shown relative to the model 3.08\,$\mu$m water band depth in the figure. Clearly, the line ratios of all
ice bands relative to the water band change dramatically with inclination. Note that this is
not an effect of varying ice abundances within the disk, since both H$_2$O and CO$_2$ ices are assumed to evaporate at 90\,K. 
The effect is especially apparent for the 3.08\,$\mu$m and 6.0\,$\mu$m water ice bands, the ratio of which drops from $\sim10$ at an inclination of $65\degr$ to $\sim0.7$ at an inclination of $74\degr$.
At even higher inclination, the ratio between the two water ice bands drops to a plateau of 0.2-0.25.

Note that the varying ratios of the ice absorption bands are, in this case, a purely 2-dimensional effect. The `classical' radiative transfer scenario for ice bands is material in a cold slab seen toward a background
infrared point source. The important difference for observing ices in a disk is that the
background source is extended relative to the absorbing material. For instance, if the infrared source is due to scattered light, the size of the background source may be comparable to the size of the flared disk. There are two mechanisms which produce the extended 
source in the case of a disk. The first is light from the central star and inner rim scattered in the surface layer of the disk, and the second is thermal emission from the surface layer. Inspection of Fig. \ref{SED_MODEL_NOFRONT_incl} illustrates the relative importance of these two effects. For the
spectrum produced by the highest inclination, a minimum in the flux is observed at $\sim8\,\mu$m. 
At shorter wavelengths, scattered light dominates the spectrum, while thermal emission from the disk surface layer dominates at longer wavelengths. This means that the 15.2\,$\mu$m CO$_2$ bending mode is always observed against a thermal source, while the rest of the main ice bands are observed toward a scattered source for a highly inclined disk. At an inclination similar to the disk opening angle, the emission from the central star and inner rim start to dominate. In this case the entire length of the disk is probed and the deepest ice bands are produced. In Fig. \ref{SED_MODEL_NOFRONT_incl}, it is seen
that the inclination where the peak absorption of the first three ice bands (the 3.08\,$\mu$m and 6.0\,$\mu$m water ice bands and the 4.27\,$\mu$m CO$_2$ ice band) occurs, increases with increasing wavelength of the band. This is, however, not true for the 15.2\,$\mu$m CO$_2$ ice band, which peaks at a lower inclination than the 6\,$\mu$m band. This is due to the fact that the first three bands are seen toward
a scattering source, while the 15.2\,$\mu$m band is seen toward a thermal source whose geometry is different.

Obviously, such behaviour significantly complicates the determination of accurate ice abundances in disks and other multi-dimensional scenarios. On the other hand, strongly deviating
ice band ratios seen toward an edge-on disk is a strong signature that the ices are intrinsic to the disk.
For the case of CRBR 2422.8-3423, no strongly deviant line ratios are observed, at least not when
compared to other ice sources. In particular, the ratio between the 6.0\,$\mu$m 
and the 3.08\,$\mu$m water ice bands is typical of embedded sources. It is a long-standing problem that the water ice band 6.0/3.08 ratio is up to two times higher than expected. Part of this may be explained by radiative transfer effects analogous to those identified here for nearly edge-on disks. However, since most observed icy lines of sight exhibit this anomalous ice band ratio, the effect is unlikely to be entirely due
to the presence of disk geometry. The CO$_2$ bending
mode of CRBR 2422.8-3423 can, unfortunately, not be compared to the stretching mode at 4.27\,$\mu$m, which is located outside the spectral coverage of Spitzer-IRS. The ratio of the two CO$_2$ ice bands could otherwise constitute an excellent tracer of radiative transfer effects in disks. 

The fact that CRBR 2422.8-3423 has an inclination which is exactly optimal for observing ices in the
disk is an interesting coincidence. Had the inclination been a few degrees higher, the mid-infrared continuum flux level would have dropped by an order of magnitude, and the ice band ratios would have deviated strongly from the norm. 
Another important conclusion is that pure CO ice is not expected to show up in spectra of disks with radii smaller than 100--150\,AU and surrounding stars of luminosity $\gtrsim 1\,L_{\odot}$. Any CO ice present should show clear signs of being embedded in a less volatile ice matrix, such as a dominance of
the red water-rich component at 2136\,cm$^{-1}$.

\section{Conclusions}
We have presented a full mid-infrared spectrum of the inclined disk CRBR 2242.8-3423 covering all of the strongest absorption features due to interstellar ices, except the 4.27\,$\mu$m CO$_2$ band. By fitting an axisymmetric, but effectively 3D, Monte Carlo radiative transfer model to the spectrum as well as photometric data covering 1.6-3000\,$\mu$m and spatially resolved near-infrared images, we have been able to constrain the location of ices in the disk and in the foreground material along this line-of-sight. The main conclusions
can be summarized as follows:

\begin{itemize}

\item The use of an axisymmetric continuum radiative transfer model to simultaneously fit a high spectral resolution SED
as well as high spatial resolution imaging is a powerful tool to estimate the physical conditions of dust
within circumstellar disks. Combining the model with high resolution temperature-dependent opacities further allows a detailed study of the icy material present in the disk.

\item The extinction of the near-infrared emission together with submillimeter imaging indicate that a dense and cold foreground component, containing a significant part of the observed ices, is present. Specifically, the line-of-sight toward CRBR 2422.8-3423 passes through the dense core Oph-F.

\item  If a constant ice abundance for all ice species apart from CO is assumed throughout the disk and foreground material for temperatures below 90\,K, up to 50\% of the water and CO$_2$ ices reside in the disk. The 6.85\,$\mu$m feature shows signatures of strong thermal processing, indicating that a similar fraction of the carrier, which is possibly NH$_4^+$, also resides in the disk. The shape of the 6.85\,$\mu$m band is arguably the strongest evidence for the presence of ices in the disk.

\item None of the pure CO ice observed toward CRBR 2422.8-3423 can be located in the model disk because the temperature is too high. The presence of a colder self-shadowed disk at larger radii is not ruled out, but CO ice in a flat outer disk will not contribute to the observed 4.67 $\mu$m band. This is because an outer disk with a component along the line-of-sight to the central star will also appear in scattered light. According to the disk model, only dust with temperatures of 40--90\,K and associated gas densities of $10^6-10^7\,$cm$^{-3}$ can contribute to the observed ice bands. Up to 20\% of the CO ice is trapped in water ice and can survive in the disk together with an additional 20\% trapped in CO$_2$ ice. For a constant abundance of these types of CO ice throughout the disk and envelope, at most 20\% of the total observed amount of solid CO originates in the disk. In general, pure CO ice
should only be observed in large quantities in disks surrounding low-mass stars if the disks have radii much larger than $\sim 150$\,AU. Indeed,  \cite{Thi02} concluded that most of the CO ice is located in the disk, based on a larger model disk. Excluding the presence of pure CO ice in the disk required the detailed 2D modeling presented here. It should be emphasized that the new Spitzer observations
indicate that warm rather than cold ices are indeed present in the disk.

\item In general, the high optical depths through the mid-planes of nearly edge-on disks prevent cold ices in the disk ($\lesssim 30$\,K) at radii of $\lesssim 200\,$AU from being directly observed in absorption toward a central 1\,L$_{\odot}$ star. 

\item The model does not provide a good fit to the red wing of the 9.7\,$\mu$m silicate band. The discrepancy is due to the absence of the strong librational band of water ice centered at 12-13\,$\mu$m. We are unable to explain the absence of the band since laboratory experiments indicate that the strength does not change appreciably under interstellar conditions.

\item The infrared source against which ice bands are observed is extended for an edge-on disk due 
to scattering at wavelengths shorter than $\sim 8\,\mu$m and due to thermal emission from the 
disk surface layer at longer wavelengths. This results in ice band ratios which are not equal to
the corresponding band ratios of the opacity. The model indicates that the ratio between the 3.08\,$\mu$m and the 6.0\,$\mu$m water ice bands
as well as the ratio between the 4.27\,$\mu$m and the 15.2\,$\mu$m CO$_2$ ice bands are excellent tracers of radiative transfer effects on ice bands intrinsic to an edge-on disk. Ratios between bands of
the same ice species are potentially powerful probes of the distribution of ices in disks. At the same time, anomalous ratios may be used to determine if the ices are located in the disk as opposed to a foreground cloud. The opposite is not true, i.e. normal band ratios do not indicate that the ices are located outside the disk.

\end{itemize}

\acknowledgments
      Support for this work, part of the Spitzer Space Telescope Legacy Science Program, was provided by NASA
      through Contract Numbers 1224608 and 1230780 issued by the Jet Propulsion Laboratory, California Institute
      of Technology under NASA contract 1407. KMP was supported by a PhD grant from the Netherlands Research School for Astronomy (NOVA). Astrochemistry in Leiden is supported by a SPINOZA graint of the Netherlands Organization of Scientific Research (NWO). 
      The authors are grateful to Doug Johnstone for
      providing the SCUBA 850\,$\mu$m map from the COMPLETE survey and Karl Stapelfeldt for comments which improved the manuscript. Finally, we wish to acknowledge the efforts of all the people who put many years of hard work into making the Spitzer Space Telescope IRS instrument a reality.

\bibliographystyle{apj}
\bibliography{ms}

\end{document}